\documentclass[]{cas-dc}

\usepackage{l3regex}

\xspaceaddexceptions{]}
\def\tsc#1{\csdef{#1}{\textsc{\lowercase{#1}}\xspace}}
\tsc{WGM}
\tsc{QE}
\tsc{PMS}
\tsc{BEC}
\tsc{DE}

\usepackage{float}
\usepackage{amsmath}
\usepackage{nccmath}

\usepackage{adjustbox}
\usepackage{multirow}

\usepackage{multicol}
\usepackage{caption}
\usepackage{hyperref}

\usepackage{natbib}
\begin{document}

	\let\WriteBookmarks\relax
	\def\floatpagepagefraction{1}
	\def\textpagefraction{.001}
	\shorttitle{D. Godines et al. / Lens Identification Algorithm}
	\shortauthors{D. Godines et al.}
	
	\title [mode = title]{A Machine Learning Classifier for Microlensing in Wide-Field Surveys}                      
	\author[1]{D. Godines}[orcid=0000-0001-8495-8205]
	\cormark[1]
    \ead{danielgodinez123@gmail.com}
	\author[1]{E. Bachelet}
	\author[2]{G. Narayan}
	\author[1]{R.A. Street}
	
	\address[1]{Las Cumbres Observatory, 6740 Cortona Drive, Suite 102, Goleta, CA, 93117, USA.}
	\address[2]{Space Telescope Science Institute, 3700 San Martin Drive, Baltimore, MD, 21218, USA.}
	
	\cortext[cor1]{Corresponding author}
	
	\begin{abstract}
		While microlensing is very rare, occurring on average once per million stars observed, current and near-future surveys are coming online with the capability of providing photometry of almost the entire visible sky to depths up to R $\sim$22 mag or fainter every few days, which will contribute to the detection of black holes and exoplanets through follow-up observations of microlensing events. Based on galactic models, we can expect microlensing events across a vastly wider region of the galaxy, although the cadence of these surveys (2-3 $d^{-1}$) is lower than traditional microlensing surveys, making efficient detection a challenge. Rapid advances are being made in the utility of time-series data to detect and classify transient events in real-time using very high data-rate surveys, but limited work has been published regarding the detection of microlensing events, particularly for when the data streams are of relatively low-cadence. In this research, we explore the utility of a Random Forest algorithm for identifying microlensing signals using time-series data, with the goal of creating an efficient machine learning classifier that can be applied to search for microlensing in wide-field surveys even with low-cadence data. We have applied and optimized our classifier using the OGLE-II microlensing dataset, in addition to testing with PTF/iPTF survey data and the currently operating ZTF, which applies the same data handling infrastructure that is envisioned for the upcoming LSST.
	\end{abstract}
	
	\begin{keywords}
		 gravitational microlensing \sep classification \sep 
		 \sep random forest  
		 \sep machine learning
		 \sep PTF
	     \sep ZTF
	\end{keywords}
	\maketitle
	
\section*{1. Introduction}
\label{intro}
Gravitational microlensing allows for the study of numerous objects that may otherwise be too faint to observe. This is done by observing the alignment of a foreground object along the observer's line of sight to a luminous background source, a method which even allows for the study of otherwise invisible material such as black holes \citep{wyrzykowski2016black}. Over the last several decades several surveys have successfully used microlensing to study other phenomena such as compact objects \citep{griest1991galactic}, and exoplanets orbiting foreground lens stars \citep{microlensing_planet}.
	
One of the teams leading the microlensing effort is OGLE \citep{udalski1992optical}, which based at the Las Campanas Observatory, Chile, is currently in the fourth phase of its survey (OGLE-IV), covering 3000 square degrees in the most crowded regions of the sky, including the Galactic Bulge and Magellanic System \citep{2015OGLE_4}. While the MACHO \citep{alcock2000macho} and EROS \citep{ansari2004eros} teams helped lead the microlensing efforts, current microlensing surveys are being led by teams such as MOA \citep{bond2001real}, KMTNet \citep{kim2016kmtnet}, as well as Gaia \citep{Hodgkin2013} and the All Sky Automated Survey for SuperNovae (ASAS-SN, \citet{Shappee2014}, \citet{Kochanek2017}).

These surveys search for microlensing by employing time-series imaging to detect any increase in brightness that may result from the magnification of the source star during lens-source alignment. These events occur on timescales ranging from $\sim$1 day, to up to hundreds of days depending on the mass, distance and relative motion of the lens and source objects. The unpredictable nature of microlensing thus requires data to be gathered at high cadence ($<$1 hour to one day) for optimal results, as short-lived anomalies that last minutes to days could signal the presence of companion objects present in the lensing system. As rare transient events, it's important to detect microlensing as early as possible so as to be able to monitor any anomalies that may occur, and as such it is imperative to detect these events from a real-time data stream. Detecting microlensing during the rising phase of the lightcurve allows us to optimize observing strategies to ensure adequate data can be gathered to fully characterize the event, which is particularly necessary to constrain the rapid change in flux that transpires during anomalies. 
	
The detection and classification of astrophysical phenomena from time-series data is currently receiving renewed attention owing to recent and anticipated developments in survey technology. Medium ($\sim$1-2m) and large-aperture ($\sim$4m+) telescopes with wide-fields of view can now be equipped with high spatial resolution detectors to photometrically survey hundreds of square degrees per night, with examples including the 1.2m Zwicky Transient Facility (ZTF, \cite{BellmZTF}), the 4m VISTA survey telescope \citep{VISTA_2015}, as well as the upcoming 8.4m Large Synoptic Survey Telescope (LSST, \cite{tyson2002_lsst}). These new surveys are employing improved software designed to make it possible to issue real-time alerts in a timely manner, allowing for the follow-up of transient phenomena (\citet{Law2009}, e.g. PTF). Even though microlensing is an intrinsically rare phenomenon, with the optical depth toward the Bulge at $\approx10^{-6}$ (\citet{alcock2000macho}; \citet{sumi2016possible}), the large field of view of these surveys could dramatically broaden the region of the galaxy over which microlensing events are detected, enabling us to study populations in different evolutionary contexts. 
	
The main surveys of ZTF and LSST can and will produce low-cadence data ($\sim$1-3 $d^{-1}$) rather than multiple measurements per night, making microlensing harder to detect as these signatures have correspondingly lower signal to noise particularly during the rising phase. It is therefore timely to reevaluate the detection algorithms used to filter for these transient phenomena. Previous authors have outlined different methods for microlensing detection in wide-field surveys, such as the OGLE team which have employed their Early Warning System (EWS) designed to detect on-going microlensing events, which continues to be employed and aided in the identification of 40-80 microlensing events per year during the OGLE-II phase \citep{udalski2003optical}. Similar algorithms were put into effect by the MOA collaboration \citep{bond2001real} as well as the MACHO group \citep{alcock1997macho}. While the EWS system, for example, is capable of analyzing photometry in real-time to detect ongoing microlensing events from high cadence data, the Korea Microlensing Telescope Network (KMTNet) applies an event-finding algorithm that performs linear fits to a grid of point-lens microlensing models and in turn works best with complete lightcurves of the event \citep{kim2018korea}. 
	
Real-time detection of microlensing in low-cadence data is a tremendous challenge as the timescale and amplitude of the event can vary significantly depending on the parameters of the lens-source alignment. \citet{price2014statistical} searched for microlensing in the irregularly-sampled PTF database by filtering for potential candidates via lightcurve statistics, and while he reported three plausible signals he could not confirm with certainty due to the gaps in the photometry. Machine-learning techniques have shown great promise when it comes to lightcurve classification of low-cadence data. \citet{Richards2011} demonstrated the utility of machine-learning in his application of ten different machine-learning engines to classify noisy and sparse lightcurves, reporting misclassification rates between 23 and 32 percent with the highest accuracy yielded by the Random Forest algorithm. 

This research aims to further explore the application of the Random Forest algorithm to detect microlensing events in real-time, with both high cadence and low cadence survey data, with the intention of developing an open-source program that can be applied with ease by any member of the astronomy community in the search for microlensing. Section~\hyperref[psplmodel]{2} describes the microlensing theory followed by a section detailing the Random Forest ensemble learning method. The development of a proper training set, including the simulation of lightcurves and the features extracted from them to train the algorithm are described in Sections~\hyperref[trainingset]{4} and \hyperref[featureselection]{5}, respectively. In addition, Sections~\hyperref[hyperparameteroptimization]{5.2} to \hyperref[ZTF18abegegr]{8.3} illustrates the performance of this machine-learning algorithm when applied to real data from several surveys. We conclude in Section~\hyperref[conclusion]{9} with a presentation of our classifier, Lens Identification Algorithm (LIA), as well as our plans to integrate this code into the ANTARES touchstone which will serve as a broker for the Large Synoptic Space Telescope (LSST, \citet{Gautham_Antares}).
	
\section*{2. Microlensing Model}
\label{psplmodel}
For our purposes of simulating microlensing lightcurves with which to train a supervised machine learning engine, we sought to simulate only single-lens microlensing events (single point-lens \& point source) which can be described by three parameters, the Einstein crossing time ($t_E$), defined as the time to cross the lens' Einstein radius, $R_E$, the minimum impact parameter ($u_0$), and the event peak time ($t_0$). With these parameters one can define the event lightcurve with the following three functions of time: its amplification factor $A(t)$, which describes the magnification of the event, the observed flux as a function of time $F(t)$, and the normalized angular distance between the source and the lens, $u(t)$, defined as
\begin{ceqn}
\begin{align}	
u(t) = \sqrt{u_{0}^2+\left(\frac{t -t_0}{t_E}\right)^2} \,,
\end{align}
\end{ceqn}
\begin{ceqn}
\begin{align}	
A(t) = \frac{u^2+2}{u\sqrt{u^2+4}} \,,
\end{align}
\end{ceqn}
\begin{ceqn}
\begin{align}	
F(t) = A(t) \times f_s,
\end{align}
\end{ceqn}
where $f_s$ is the source flux \citep{paczynski1986gravitational}. This assumes that the source flux in the CCD frame is isolated, such that the flux can be measured independently of any stellar neighbors. Unfortunately the most promising regions for microlensing detection, the Bulge and the Magellanic Clouds, are extremely crowded and the blending of light can yield inaccurate measurements for $A(t)$ \citep{han1999analytic}. To account for this blending, $F(t)$ is calculated as 
\begin{ceqn}
\begin{align}	
F(t) = A(t) f_{s} + f_b,
\end{align}
\end{ceqn}
where $f_b$ is the blend flux. The overall observed flux is then
\begin{ceqn}
\begin{align}	
A_{obs}(t) = \frac{f_sA(t)+f_b}{f_s+f_b}.
\end{align}
\end{ceqn}
Taking $g = \frac{f_b}{f_s}$, $A_{obs}(t)$ can be expressed as 
\begin{ceqn}
\begin{align}	
A_{obs}(t) = \frac{A(t)+g}{1 + g} \,.
\end{align}
\end{ceqn}
Ultimately accounting for blending requires guessing initial event parameters to derive an initial model for A(t), after which the values for $f_b$ and $f_s$ can be inferred through a model-fitting process. While constraining $f_b$ through the fitting process is the most common method for dealing with blending, it is sometimes possible to actually resolve the stars contributing to $f_b$ through the use of space or large ground-based telescopes \citep{janczak2010sub,bennett2006identification}. For our purposes of detecting these events, we simulate these signals by setting only a value for the blending coefficient $g$, derived from a reasonable distribution (see Section~\hyperref[microlensingmodel]{4.3}). While in reality a star is not a point source and a more complex model can be applied to account for additional behavior such as binary and parallax effects, this simple model is sufficient for our purposes of classifying microlensing events in real-time to properly follow-up any anomalies.
	
\section*{3. Classification via Machine Learning}
\label{machine_learning}
The task of quick, automated classification in astronomy has been tackled successfully over the past decade through the use of machine learning, both as a means of detecting particular classes of stars as well as for identifying instrumental artifacts. The Palomar-Quest (PQ) survey operated by the Palomar-Quest Consortium circumnavigated the problem of artificial artifacts (saturation, instrument glitches, etc.) by implementing an Artifical Neural Network (ANN) based classifier \citep{ripley1996pattern} into their data reduction pipeline \citep{djorgovski2008palomar}, which in turn limited transient false-alerts. This illustrates the utility of morphological image processing, which when coupled with other machine learning methods can be applied as an effective method for maintaining catalogs as well as performing automated image classification \citep{Weir1995, Odewahn2004}. ANN was also applied by \citet{OGLE_MachineLEarning2003} to filter for eclipsing binaries among variable stars in the OGLE-II catalog, while the search for and characterization of microlensing events in the OGLE-III database was facilitated using the Random Forest algorithm (\citet{2015_OGLE_RF}; \citet{2016_OGLE_RF}). While there are other machine learning methods that have been utilized with great success, such as Support Vector Machines (SVMs) \citep{cristianini2000introduction} employed for example by the Nearby Supernova Factory based in the Lawrence Berkeley National Laboratory \citep{romano2006supernova}, a side-by-side comparison of ten different classifiers in two different datasets (OGLE and Hipparcos) by \citet{Richards2011} found the Random Forest to have the lowest error rate. 
	
The Random Forest is popular for classification tasks as a result of its accuracy, speed, and straightforward application; and when tested against other machine learning algorithms tends to be among the top performers. In addition to the study performed by \citet{Richards2011}, \citet{Pashchenko2018} recently reported the Random Forest as a top performing machine learning algorithm when tasked with classifying variable stars, performing equally as well as SVM, gradient boosting and neural networks, with logistic regression and k-nearest neighbors algorithms displaying significantly lower performance in this particular study. Given the successful application of the Random Forest algorithm for classification tasks, we seek to apply it for our purpose of differentiating between microlensing signals and other transients and variables. 
	
\subsection*{3.1 Random Forest Algorithm}
\label{random_forest}
Classification via supervised machine learning is performed by using a training set containing sources of known class, with each being described by a set of input variables. The goal of a machine learning classifier is to create a function that can map the input variables to the output class, which in turn can be employed to predict the class membership of unknown objects, creating an automated process which requires no human input and, when properly trained and validated, can provide a high performing classification algorithm that can be used to quickly categorize new objects with a statistically low error rate \citep{Brink2013}.
	
The Random Forest is an ensemble machine learning method developed by \citet{Breiman2001} that trains numerous decision tree classifiers and takes the mode of the classification results as the output, in effect combining a multitude of ``weak'' learners (individual decision trees), to create one ``strong'' learner (the ensemble). Decision tree learning is used to map a set of input features to output classes by means of a series of selection rules determined during the training process, and individual trees, when trained deep enough, can create complex structures given the feature space; although a single tree has high variance and being sensitive to noise, it alone is not sufficient for wide-scale classification tasks. In the Random Forest ensemble, each tree is trained with a subsample of input features, and as such the entire ensemble is in turn de-correlated yielding low overall variance, resulting in a high performing classifier that will run the input features through the individual trees, providing a prediction based on the most frequent class that is output once all the trees have ``voted''. For this research we used the open source machine learning software package \textit{scikit-learn}, implemented in Python \citep{scikit-learn}. 
	
This implementation is based on bootstrap aggregation, or \textit{bagging} \citep{breiman1996bagging}, in which $k$ subsamples ($B_i$) are created from the training data $B$, with each tuple in $B$ represented by some $n \times 1$ vector containing information on $n$ features. Each $B_i$ is constructed with replacement at each iteration, such that some values in $B$ may not be present in any of the bootstrap samples, whereas others may be repeated across several $B_i$. The data in each subsample is then used for training by creating $k$ decision trees, using the \textit{scikit-learn} classification and regression trees (CART\footnote{For a detailed description, see: http://scikit-learn.org/stable/}) algorithm, with the out-of-bag data that was not included in the bootstrap used for testing. During the tree-building process, subsequent splits at each node are chosen amongst all the available features, with optimal split points deducted using the Gini-index \citep{breiman1984classification}, which serves as a means of measuring impurity in $B$. The Gini-index is defined as
\begin{ceqn}
\begin{align}	
\text{Gini index} = 1 - \sum_{i=1}^{n} p_{i}^2
\end{align}
\end{ceqn}
where $p_i$ is the probability that a given tuple in the training data $B$ belongs to the `$i^{th}$' class. A low Gini-index indicates that $B$ was well partitioned, serving as a means of selecting the best splits across all nodes. This bootstrap aggregation method reduces the overall bias of the final model as the ensemble is able to train with numerous subsamples of the data, whereas simply training with half of the data and testing with the other half may yield lower performance as some data used for testing may differ significantly from that of the training set. At the end, each tree is constructed using a different bootstrap sample of the data, and classification of new data is performed by running the features through the trained decision trees, with the class that is output the most in the ensemble taken to be the final prediction. In this manner, each prediction is accompanied by probability predictions, which is the normalized quantity of individual trees that yield each individual class. Thus, if a similar number of trees vote for several classes, their probability output will be similar indicating confusion within the classification engine for the particular data. By enforcing a threshold on this metric, for example, requiring a probability prediction of 0.5 before accepting the classification as true, one can limit the number of false-alerts at the risk of disregarding true-positives that for whatever reason are not well-defined in the feature space. 
	
\section*{4. Initial Training Set}
\label{trainingset}
In order to build a training set of known objects at the start of a survey, one would typically rely on data from previous surveys or from simulations, with the quantity and quality of training data having a significant impact on the performance of the classifier, as shown by \citet{Brink2013} in which he reported a higher performing classifier than that of \citet{Bloom2012} within the same dataset, ultimately attributed to the substantially larger training set provided as a result of data availability at the time of research. As real data can be sparse and not readily available, especially at the start of a survey, we chose to simulate our training set with adaptive cadence. This not only generalizes the algorithm for application across any dataset, but also ensures our training set is well-representative of the source class, as using real data to train requires a substantial quantity of quality samples that may not always be available. A training set with adaptive cadence in this context is constructed by taking real lightcurves from a given survey, and simulating the source classes given only random timestamps selected from said lightcurves. When no lightcurves are available as is the case when the survey is not yet underway, the timestamps can be derived from the planned observation strategy; either way, this method of simulating with adaptive cadence works best if one can accurately depict the intervals at which a survey will collect data. When coupled with an accurate noise model, this technique allows us to accurately simulate how a given survey would observe the particular classes we describe in the following sections. 
	
\subsection*{4.1 Cataclysmic Variables}
\label{cv_model}
We seek to simulate data from the most troublesome sources that can often times yield false-alerts when searching for microlensing events. Spontaneous eruptions by cataclysmic variables (CV), in particular dwarf novae, can mimic a microlensing lightcurve that will trigger false-alerts especially when only single-band photometric data is available. CV's are binary systems with relatively short periods, consisting of a white-dwarf accreting matter from a larger companion star via Roche Lobe overflow \citep{howell2001exploration}. These systems are known for their outbursts, which occur due to instability within the accretion disc surrounding the white dwarf resulting from accreted material falling into the star \citep{osaki1974accretion}. This mechanism is what causes the eruption that increases the flux of the system, and as the disc is drained of its matter the outburst subsides, continuing to acquire more material via Roche Lobe overflow until the same mechanism is again triggered and a new outburst occurs (for a thorough explanation, see \citet{hellier2001cataclysmic}). Although CV outbursts occur repeatedly and usually rise faster than microlensing events, this is not always the case and are thus troublesome in the search for microlensing especially when no previous baseline data is available to reject these events \citep{kim2018korea}.
	
CV's are usually classified as either novae, such as supernovae that yield a one-time high magnification eruption; recurrent novae which remain dormant for long periods of time and typically exhibit outbursts every 10 to 100 years; and dwarf-novae, which display lower-amplitude outbursts than the previous two subclasses but at higher frequency, ranging from 10-50 days for Z Camelorpardalis stars or 15-500 days for U Geminorum stars (\cite{robinson1976structure}; see his Table 1). To simulate dwarf-novae, we approximated each outburst as three straight line functions of different gradients dependent on the maximum outburst amplitude, $A_{max}$. We first generate a period, $P$, from a normal distribution with a mean of 100 days and a standard deviation of 200 days, followed by an outburst amplitude $A_{max}$ selected from a uniform random function between 0.5 and 5.0 mag. As the morphology parameters of each outburst varies \citep{hellier2001cataclysmic}, we first generate the start time of the first outburst anywhere between the initial timestamp, $t_0$, and $t_0+P$. The start time of each subsequent outburst ($t_{outburst}$) that occurs within the timestamps are computed from the end of the first outburst, such that the parameters for each outburst are simulated as follows:
\begin{figure}
		\includegraphics[width=8.4cm,height=7.0cm]{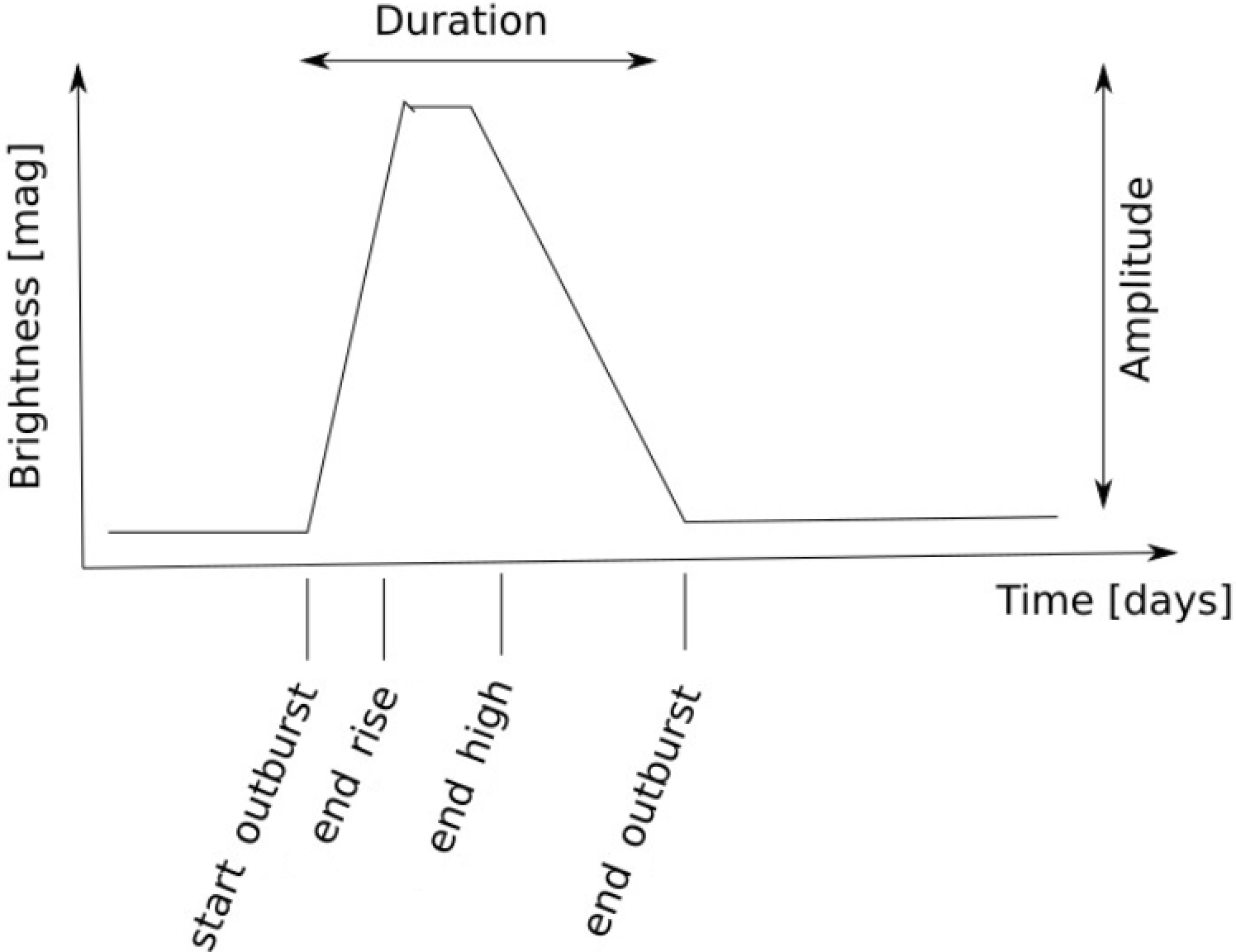}
		\caption{Dwarf novae outburst illustration. The outburst can be well represented as three linear phases, with the rise computed as a positive gradient followed by a flat phase when it reaches the maximum amplitude. The drop is a shallower negative gradient until it reaches baseline magnitude. 
		}
		\label{CV_Diagram}
\end{figure}
\begin{list}{$\circ$}{}
		\item The outburst duration, $d_{CV}$, is chosen from a uniform distribution between 3 and $\frac{P}{10}$ days.
		\item The time it takes the outburst to reach maximum amplitude, $t_{rise}$, is selected randomly from a uniform distribution between 0.5 and 1.0 days.
		\item The time the outburst remains at maximum amplitude, $t_{high}$, is selected from a normal distribution with $\mu = 0.4$ and $\sigma = 0.2$ days.
		\item The time it takes the outburst to drop back to baseline is $t_{drop} = d_{CV} - t_{rise} - t_{high}$.
		\item The time at which the outburst reaches maximum is $t_{end \ rise} = t_{start \ outburst} + t_{rise}$.
		\item The time at which the outburst begins to descent back to baseline is $t_{end \ high} = t_{start \ outburst} + t_{rise} + t_{high}$.
\end{list}
Working in magnitude scale, the rise of each outburst can be represented as a steeply positive gradient of the form
\begin{ceqn}
\begin{align}
g_{rise} = \frac{-A_{max}}{t_{rise}},
\end{align}
\end{ceqn}
followed by a flat phase at $A_{max}$, with the drop phase expressed as a negative gradient,
\begin{ceqn}
\begin{align}
g_{drop} = \frac{A_{max}}{t_{drop}}.
\end{align}
\end{ceqn}
Lastly, the magnitude $m(t)$ per each outburst during all three phases is calculated as
\begin{enumerate}
	\item $m(t) = g_{rise}\left(t - t_{start \ outburst}\right) \ | \\ \ t_{start \ outburst} \leq t\leq t_{end \ rise}$
	\item $m(t) = -A_{max} \ | \ t_{end \ rise} \leq t\leq t_{end \ high}$
	\item $m(t) = -A_{max} + \left(g_{drop}\left(t - t_{start \ outburst}\right)\right) \ | \\ \ t_{end \ high}~\leq~t\leq~t_{end \ outburst}$
\end{enumerate}
A sample schematic is displayed in Figure~\ref{CV_Diagram}, with examples of simulated lightcurves shown in Figure~\ref{CV_lightcurve}. To ensure we simulate only credible CV lightcurves, we require that there must be at least 7 measurements within any one outburst, with at least one point within the rise or drop. This is done so as to discard simulated CV that, due to poor cadence, display no CV characteristics. As per the randomness in our simulations, CVs with little to no measurements within any outburst could indeed be simulated, seemingly mimicking a constant source with several outliers. Thus, imposing these conditions ensures every simulated lightcurve has at least one adequate outburst. 
\begin{figure}
		\includegraphics[width=8.4cm,height=7.0cm]{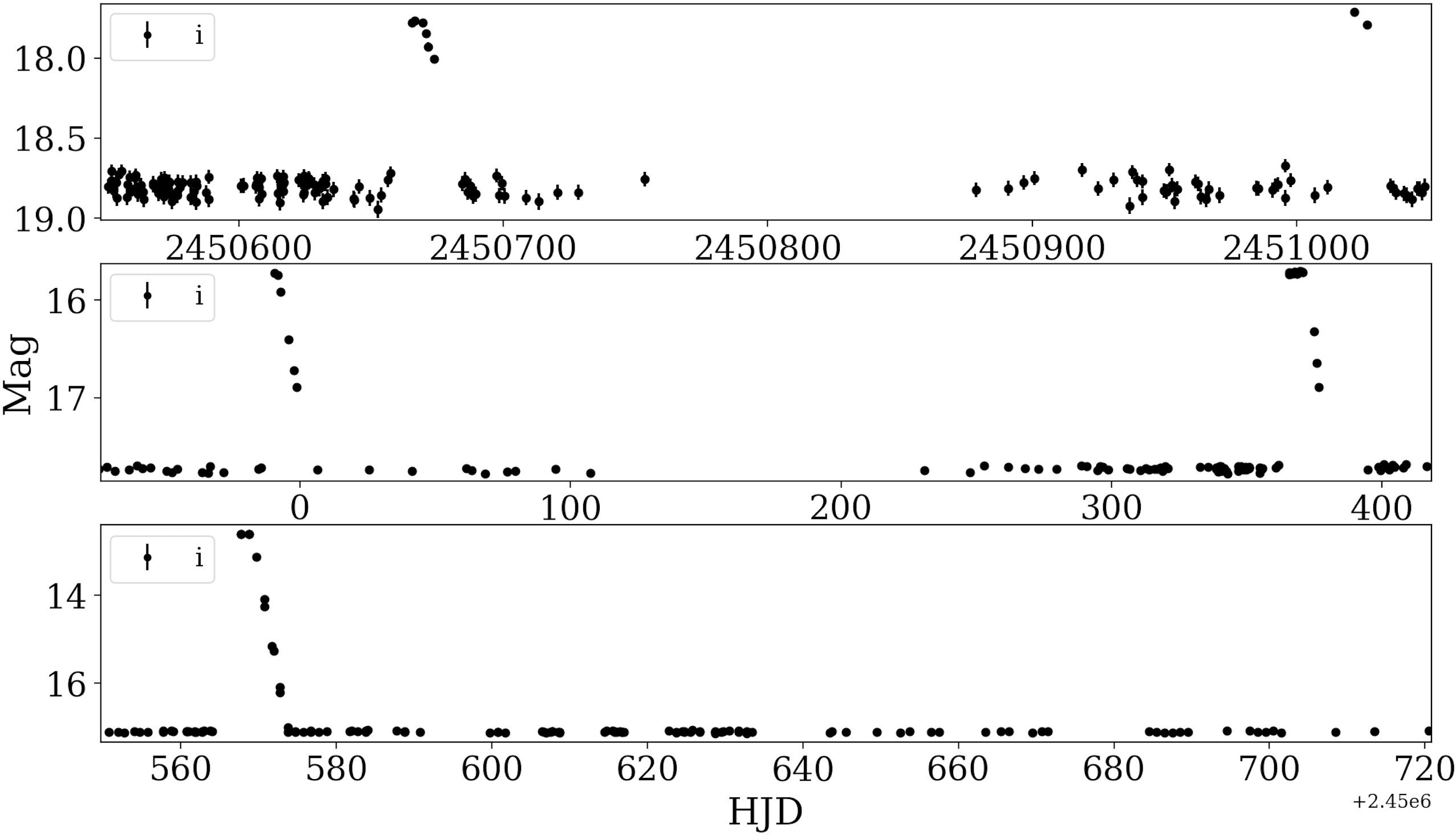}
		\caption{Example of simulated dwarf novae. Noise included using the noise model described in Section~\hyperref[iptfsearch]{7.2}.
		}
		\label{CV_lightcurve}
\end{figure}
	
\subsection*{4.2 RR Lyrae \& Cepheid Variables}
\label{variable_model}
Another class of pulsating variables that can mimic microlensing behavior are RR Lyrae, which are variable stars displaying asymmetric periodic lightcurves with periods of less than one day, pulsating as a result of mass loss that triggers oscillations within the star \citep{1998galactic_book}. Similar stars of higher mass but with longer periods are categorized as Cepheid variables \citep{1998galactic_book}, and while these stars pulsate periodically unlike the sporadic nature of CVs, they can nonetheless yield false alerts in the search for microlensing if there's no baseline data available, and especially if the pulsation is detected near the rise or fall. It is also difficult to distinguish between microlensing events and longer-period variables, and while these may be ruled out by putting limits to the RMS of the flux, it's not a sure way of excluding variability as some of these sources may vary more or less in a given year, or remain dormant and suddenly pulsate, resulting in potential false-alerts \citep{kim2018korea}. 

\begin{figure}
		\includegraphics[width=8.4cm,height=7cm]{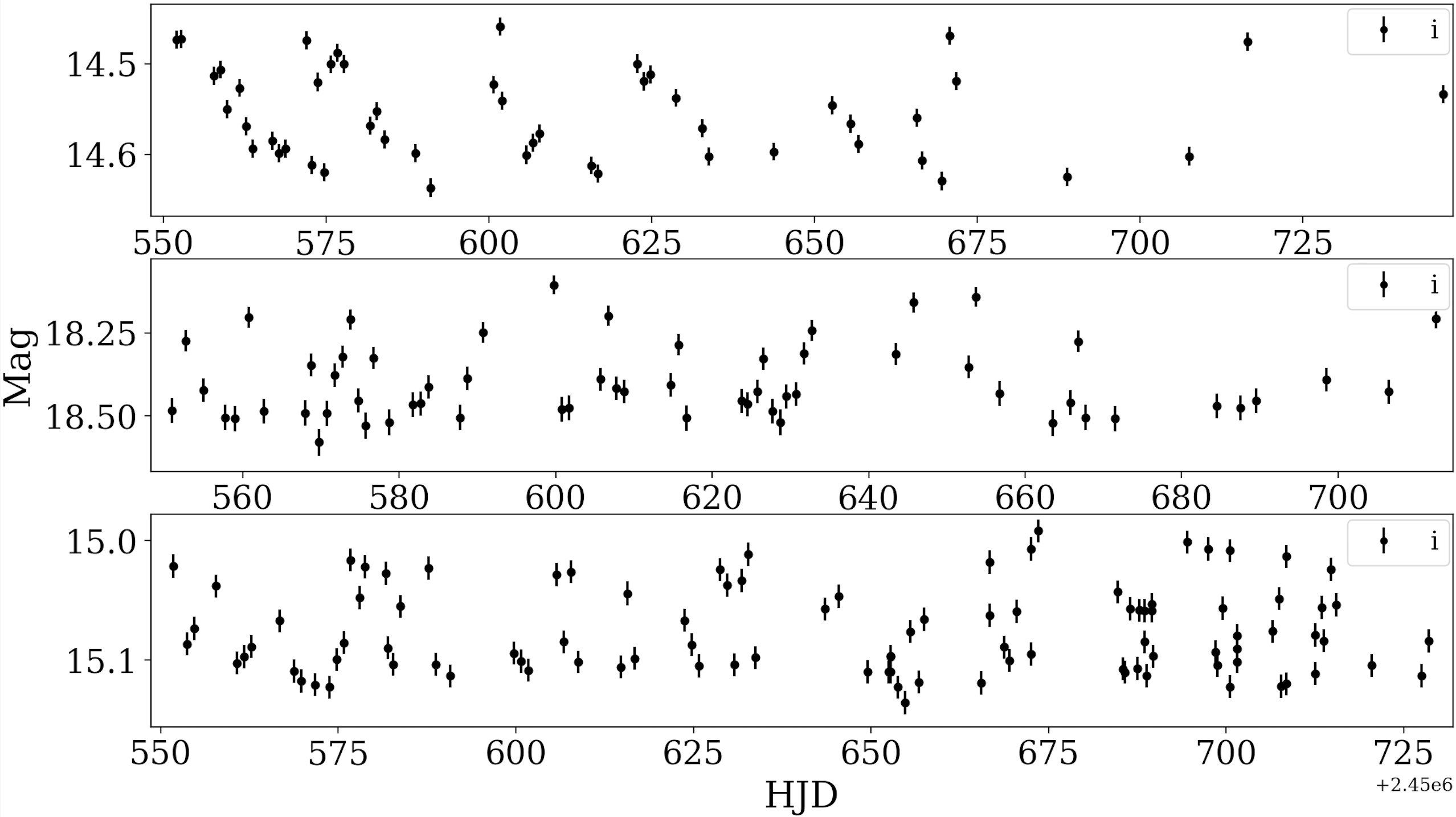}
		\caption{Example of simulated variables. Noise included using the noise model described in Section~\hyperref[iptfsearch]{7.2}.
		}
		\label{var_lightcurve}
\end{figure}

We chose to include only RR Lyrae and Cepheids in our variable class as these type of variables are primary sources of confusion with microlensing events detected in real-time and their respective short and longer period behavior allows us to capture the general characteristics of variable signals. Furthermore, we note that the program is currently designed to operate in conjunction with brokers like ANTARES, which already filters for common variables not included in our classes \citep{Gautham_Antares}. To simulate RR Lyrae and Cepheids we utilized \textit{gatspy}, an open-source Python package for astronomical time-series analysis (\citealp{2015gatspy,jake_vanderplas_2015_14833}). We made use of its template-based fitting method, which imports 483 real RR Lyrae sources in Stripe 82 (from SDSS) measured approximately over a decade \citep{2010Sesar}. By using using these lightcures as templates, \textit{gatspy} will fit an RR Lyrae model to the template, and once fitted allows us to assign a period to the model. Given the similarities between RR Lyrae and Cepheids, this technique allows us to also simulate Cepheids using the RR Lyrae data from \citet{2010Sesar} by simply fitting for longer periods. We split our RR Lyrae classes into Bailey types ab and c, with the periods for RR Lyrae ab extracted from a normal distribution centered about 0.6 days with a standard deviation of 0.15 days, while the periods for RR Lyrae type c being shorter and extracted from a normal distribution with a mean of 0.33 days and a standard deviation of 0.1 days, distributions approximated using data from \citet{2010Sesar}. In order to simulate Cepheids we derived periods from a log-normal distribution centered around 0 with a spread of 0.2 (allowing only positive periods), estimated from \citet{1977Becker} (see his Figure 7). While \textit{gatspy} allows for multi-band template fitting, we fit only the 'r' band data, as our goal is to develop an algorithm that can work when only single-band data is available. Figure~\ref{var_lightcurve} displays examples of these simulated lightcurves. 
\subsection*{4.3 Microlensing Events}
\label{microlensingmodel}
To simulate a PSPL event we first start by simulating the source, which can be taken to be a star displaying no variability, hereby referred to as a constant star. In the absence of noise, these are stars with a baseline magnitude that can be precisely replicated at every timestamp. Once noise from an appropriate model is inserted, we construct $A_{obs}(t)$ as defined in Section~\hyperref[psplmodel]{2} by referring to an analysis of microlensing events making use of six years of OGLE-III microlensing observations compiled by \citet{tsapras2016ogle}. Based on the microlensing parameter distributions for the non-blended events presented in this study, we decided on the following values for our PSPL parameters:
\begin{figure}
		\includegraphics[width=8.4cm,height=7.0cm]{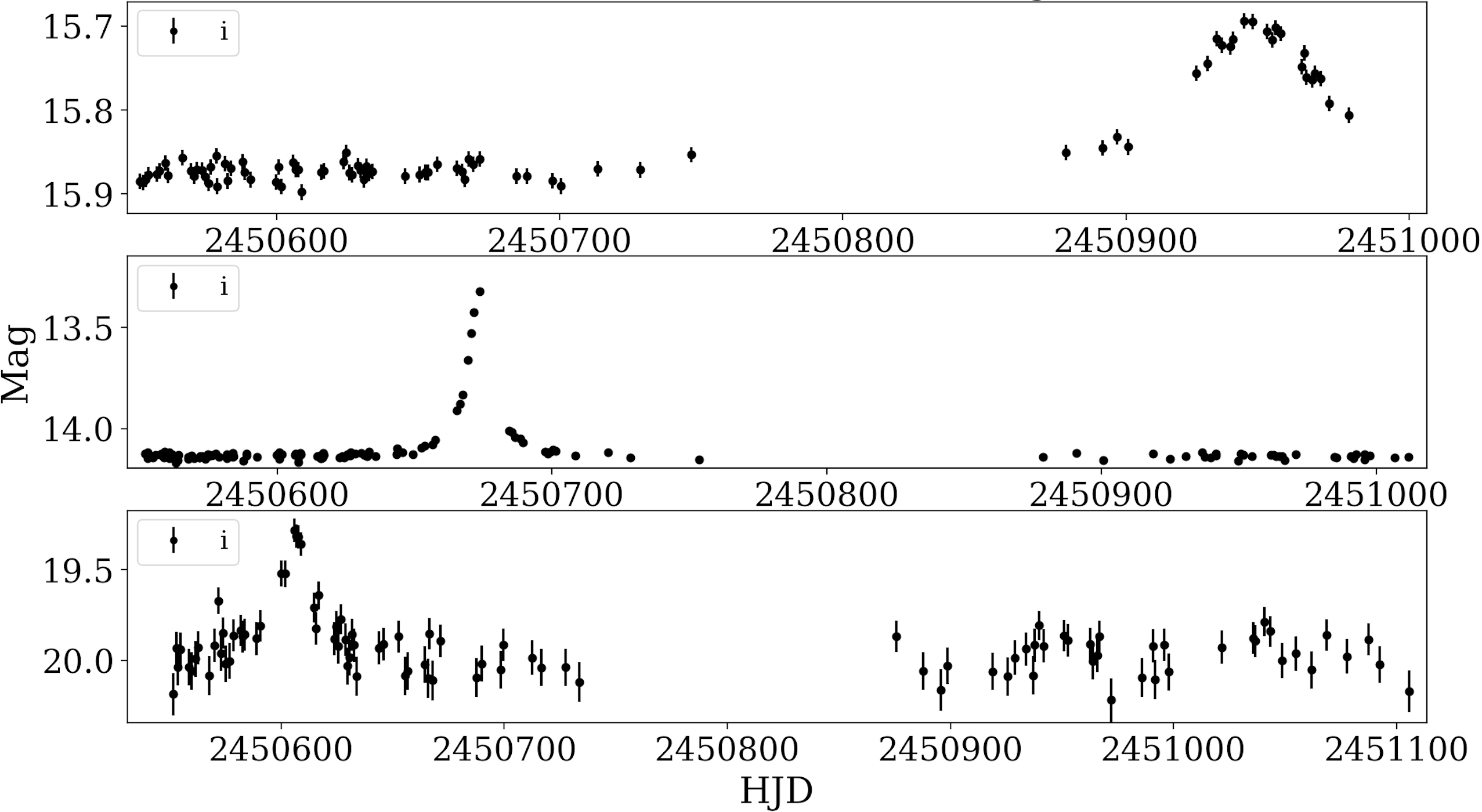}
		\caption{Example of simulated microlensing events. Noise included using the noise model described in Section~\hyperref[iptfsearch]{7.2}.
		}
		\label{ML_lightcurve}
\end{figure}
\begin{enumerate}
	\item The time of maximum amplification $t_0$ is randomly selected to be between the 10th and 90th percentile of the timestamps to ensure adequate measurements can be extracted around the peak.
	\item The angular impact parameter $u_0$ is chosen from a uniform random distribution between 0 and 1.0.
	\item The Einstein crossing time $t_E$ is chosen from a normal distribution with a mean of 30 days and a standard deviation of 10 days.
	\item The blending coefficient $g$ is chosen randomly between 1 and 10.
\end{enumerate}
Given the randomness in the microlensing simulations, we had to include additional magnification thresholds to ensure a proper microlensing training set, as simulated low-magnification events with poor signal to noise are not representative of feasibly detectable microlensing lightcurves; and as we seek to ultimately follow up these events in real-time we imposed the following thresholds to ensure we simulated and trained our algorithm with credible microlensing lightcurves. 
\begin{enumerate}
	\item There must be at least 7 points within $t_0 \pm t_E$.
	\\
	\item The mean simulated magnitude between $t_0 \pm t_E$ must be brighter than the non-magnified magnitude by at least 0.05.
	\\
	\item At least one third of photometric measurements must be magnified at least 3$\sigma$, such that one third of values in [A] are greater than 3, where [A] is the following list:
	\begin{ceqn}
	\begin{align}
	[\textit{A}] = \bigg[\frac{m_{i}^{\prime} - m_i}{\sigma_i}\bigg],
	\end{align}
	\end{ceqn}
	with $i$ denoting all indices within $t_0 \pm t_E$, and $m$ is the non-simulated photometric points of the lightcurve, $m\prime$ is the simulated photometric points, and $\sigma$ is the photometric error. 
	\\
	\item The peak magnification $A_{max}$ must be $>$ $g$.
\end{enumerate}
These magnification thresholds were used to ensure that proper signal was input into our simulated microlensing events. Example lightcurves are displayed in Figure~\ref{ML_lightcurve}.

\section*{5. Feature Selection \& Performance with Training Dataset}
\label{featureselection}
\subsection*{5.1. Feature Selection}
The goal of selecting a feature set to represent the source classes is to encode as much information as possible into a single $n \times 1$ vector, where $n$ is the total number of features. These features are metrics designed to describe all aspects of the data that may be relevant for classification, and as such these features must be derivable with both sparse and heterogeneous data, as missing data is problematic and can yield inaccurate predictions. \citet{Richards2011} explored the performance of an RF classifier by utilizing time-domain statistics as the features with which to train the algorithm. We chose a similar approach as lightcurve statistics allows us to create a general-purpose algorithm that can be applied for any survey that provides time-domain photometry, and that furthermore works with high accuracy even when few measurements are available.

\citet{Richards2011} in his study of machine learning classifiers for variable sources compiled a list of helpful lightcurve statistics, including robust metrics such as the median absolute deviation and standard deviation, as well as variability indices such as the indices J and K, first suggested by \cite{stetson1996automatic}. In addition, he computed both periodic features (extracted using the Lomb-Scargle algorithm) and non-periodic metrics, reporting the periodic features to be the most important for correctly classifying variable stars. In our work, however, we seek to apply only non-periodic metrics as we want to avoid having to compute features that are computationally intensive, and while these are extremely useful for distinguishing between different classes of variables, variability is not attributable to microlensing and for our purposes any star displaying variability can be assorted into a single class. Furthermore, we expect our algorithm to run alongside brokers such as ANTARES, which includes the identification and classification of variable sources.
	
From the 21 non-periodic features \citet{Richards2011} worked with, we made use of only nine (see Table~\ref{features_table}) as we omitted metrics that were reported with low feature importance and/or were not applicable for this research, such as the \textit{QSO} features which are quasar variability matrices. From these nine we used and adapted the \textit{median buffer range}, which is defined as the percentage of fluxes $\pm$20\% of the amplitude from the median flux, creating a sister metric which computes the same percentage using instead 10\% of the amplitude as the threshold. 
\begin{figure}
	\includegraphics[width=8.4cm,height=7.0cm]{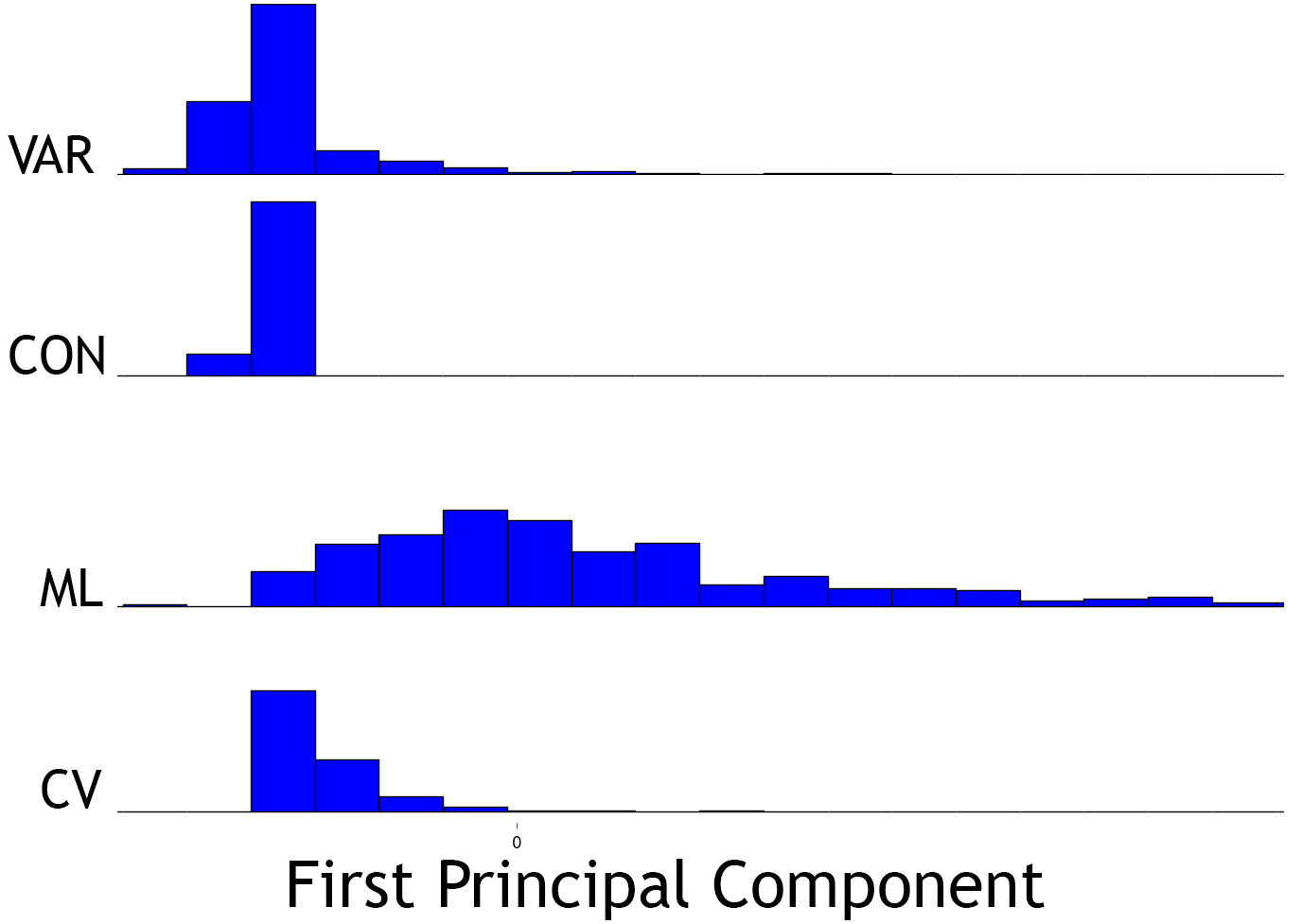}
	\caption{Histogram displaying the first principal component distribution from our principal component analysis.
	}
	\label{pca_visual}
\end{figure}
\citet{price2014statistical} also made use of the Stetson indices J and K for his work in distinguishing microlensing signals from variables, in addition to applying other variability indices compiled by \citet{shin2009detecting}, which included features such as the ratio of standard deviation to mean $\sigma / \mu$, as well as the von Neumann ratio $\eta$ \citep{von1941distribution} and $Con$, which was a measurement first used in \citet{2000Wozniak} and defined to be the number of at least three consecutive measurements that are more than 2$\sigma$ away from the median magnitude. \citet{price2014statistical} modified $Con$ to be the number of 3 or more measurements that were brighter than 3$\sigma$, as for a single-lens microlensing event one could expect this metric to yield 1, and 0 for constant sources. We've included this modified $Con$ metric in addition to this same feature but using the 2$\sigma$ threshold as originally defined by \citet{2000Wozniak}. In total we compiled 47 features, some of these which are extracted using \textit{tsfresh}, a Python package that allows for the computation of meaningful time-series features \citep{Christ2018TimeSF}. For a description on all the features used, see Table~\ref{features_table}.
\begin{figure}
	\includegraphics[width=8.4cm,height=7.0cm]{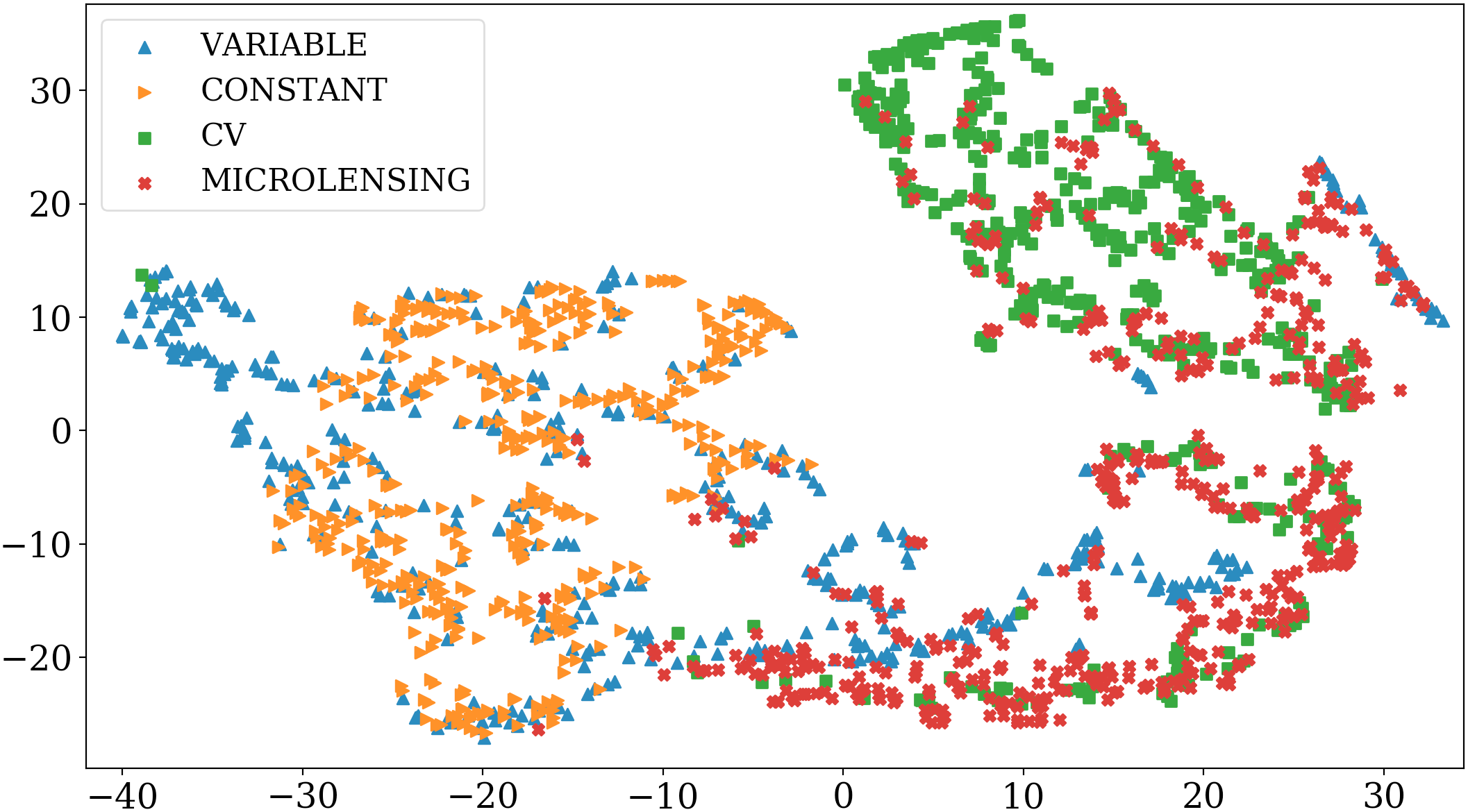}
	\caption{t-SNE projection of the 47 statistical features extracted per each lightcurve.
	}
	\label{tsne_1}
\end{figure}
\begin{figure}
	\includegraphics[width=8.4cm,height=7.0cm]{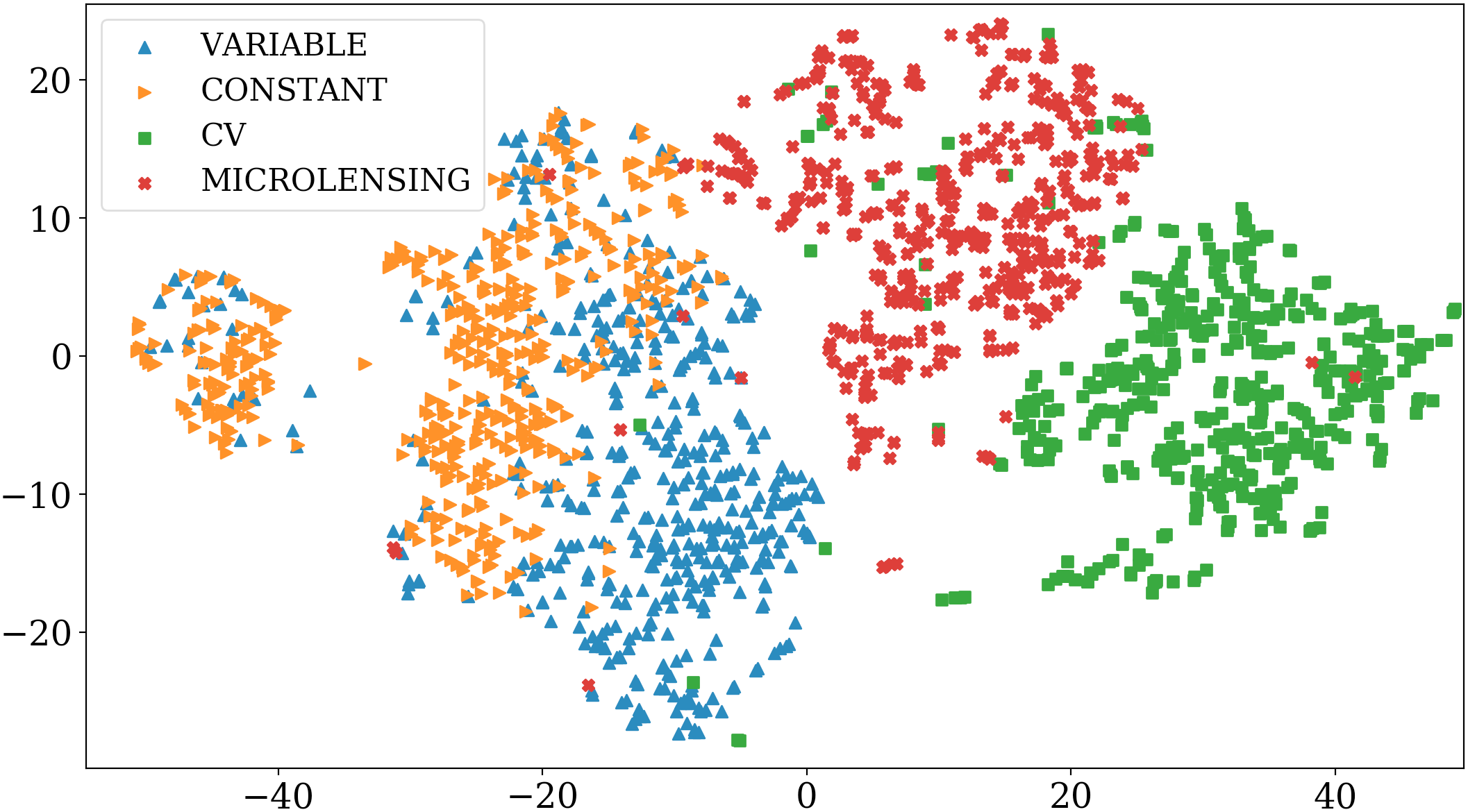}
	\caption{t-SNE projection of the principal components computed from the original feature space.
	}
	\label{tsne_2}
\end{figure}
\begin{table*}
		\centering
		\begin{adjustbox}{width=\textwidth}
			\begin{tabular}{|l|l|}
				\hline
				\textbf{Feature}&\textbf{Description} \\
				\hline
				Above 1$\ddagger$ & Ratio of data points that are above 1 standard deviation from the median. \\
				\hline
				Above 3 & Ratio of data points that are above 3 standard deviations from the median. \\
				\hline
				Above 5 & Ratio of data points that are above 5 standard deviations from the median. \\
				\hline
				Absolute Energy$^\dag$ & The sum over the squared values of the time-series. \\
				\hline 
				Absolute Sum of Changes$^\dag$ & The absolute value of the sum over the consecutive changes in the time-series. \\
				\hline 
				Amplitude$\ddagger$ & Difference between the $2^{nd}$ and $98^{th}$ percentile of the time-series. \\
				\hline
				Autocorrelation$^\dag$ & Similarity between observations as a function of a time lag between them. \\
				\hline
				Below 1 & Ratio of data points that are below 1 standard deviation from the median. \\
				\hline
				Below 3 & Ratio of data points that are below 3 standard deviation from the median. \\
				\hline
				Below 5 & Ratio of data points that are below 5 standard deviation from the median. \\
				\hline
				C3$^\dag$ & A measure of non-linearity in the time series, introduced by \citet{schreiber1997discrimination}. \\
				\hline
				Check Duplicate$^\dag$ & Checks whether any measurements in the time-series repeat at least twice. \\
				\hline
				Check Max Duplicate$^\dag$ & Checks whether the maximum value in the time-series repeats. \\
				\hline
				Check Min Duplicate$^\dag$ & Checks whether the minimum value in the time-series repeats. \\
				\hline 
				Check Max Last Loc$^\dag$ & Measures the first location of the maximum value, relative to the length of the time-series. \\
				\hline 
				Check Min Last Loc$^\dag$ & Measures the first location of the minimum value, relative to the length of the time-series. \\
				\hline
				Complexity$^\dag$ & Measured by ``stretching'' the time-series and calculating the length of the resulting line, introduced by \citet{batista2014cid}. \\
				\hline
				Con$^\mathsection$ & Number of clusters containing three or more consecutive observations larger than the baseline value plus 3 standard deviations. \\
				\hline
				Con 2 & Number of clusters containing three or more consecutive observations larger than the baseline value plus 2 standard deviations. \\
				\hline
				Count Above$^\dag$ & Number of measurements in the time-series greater than the mean value. \\
				\hline
				Count Below$^\dag$ & Number of measurements in the time-series smaller than the mean value. \\
				\hline
				First Loc Max$^\dag$ & Returns the normalized first location of the maximum value in the time-series. \\
				\hline
				First Loc Min$^\dag$ & Returns the normalized first location of the minimum value in the time-series. \\
				\hline
				Integrate & Integration of the time-series using the trapezoidal rule. \\
				\hline 
				Kurtosis$^\ddagger$ & A measure of the peakedness of the lightcurve relative to a normal distribution. \\
				\hline
				Longest Strike Above$^\dag$ & The length of the longest sequence of consecutive measurements in the time-series greater than the mean value. \\
				\hline
				Longest Strike Below$^\dag$ & The length of the longest sequence of consecutive measurements in the time-series smaller than the mean value. \\
				\hline
				Mean Absolute Change$^\dag$ & The mean over the absolute differences between subsequent measurements. \\
				\hline
				Mean Change$^\dag$ & The mean over the differences between subsequent measurements. \\
				\hline
				Mean Second Derivative$^\dag$ & The mean value of a central approximation of the second derivative. \\
				\hline
				Median Absolute Deviation$^\ddagger$ & Mean average distance between each measurement and the mean value. \\
				\hline
				Median Buffer Range$^\ddagger$ & Ratio of points that are between $\pm$ 20\% of the amplitude value over the mean. \\
				\hline
				Median Buffer Range 2 & Ratio of points that are between $\pm$ 10\% of the amplitude value over the mean. \\
				\hline
				Peak Detection & Calculates the number of peaks in the time-series. \\
				\hline 
				Ratio of Recurring Points$^\dag$ & Relative number of time-series values that appear more than once. \\
				\hline
				Root Mean Squared & The root mean square deviation of the time-series. \\
				\hline
				Sample Entropy$^\dag$ & The sample entropy of the time-series as developed by \citet{richman2000physiological}. \\
				\hline 
				Shannon Entropy & Measures the amount of information carried by a signal \citep{shannon1949mathematical}. \\
				\hline
				Skewness$^\ddagger$$^\|$ & Measures the asymmetry of the time-series. \\
				\hline
				STD$^\ddagger$ & The standard deviation of the time-series. \\
				\hline
				STD Over Mean$^\mathsection$ & Ratio of standard deviation to mean value. \\
				\hline 
				StetsonJ$^\ddagger{}^\mathsection$ & Variability index first suggested by \cite{stetson1996automatic} which measures the correlation between each measurement. \\
				\hline
				StetsonK$^\ddagger{}^\mathsection$ & Index first suggested by \cite{stetson1996automatic} which serves as a robust kurtosis measure. \\
				\hline 
				StetsonL & Variability index first suggested by \cite{stetson1996automatic} to distinguish between different types of variation. \\
				\hline
				Sum Values$^\dag$ & Sum over all time-series measurements. \\
				\hline 
				Time Reversal Asymmetry$^\dag$ & Measures the asymmetry of a series upon time-reversal \citep{schreiber2000surrogate}. \\
				\hline
				von Neumann Ratio$^\mathsection$$^\|$ & The mean square successive difference divided by the sample variance. \\
				\hline 
			\end{tabular}
		\end{adjustbox}
		\\
		$^\ddagger$ These lightcurve features were previously applied to train a machine learning classifier by \citet{Richards2011}.
		\\
		$^\mathsection$ These lightcurve features were previously applied to search for microlensing in PTF by \citet{price2014statistical}.
		\\
		$^\|$ These lightcurve features were previously applied to search for microlensing in OGLE-III by \citet{2015_Wyrzykowski}.
		\\
		$^\dag$ These metrics were computed using the \textit{tsfresh} Python package \citep{Christ2018TimeSF}.
		\\
		\captionsetup{justification=centering}{\textbf{Table 1.} All 47 statistical features extracted from the lightcurves and used for classification.}
		
		\captionlistentry{}
        \label{features_table}
\end{table*}

Given our 47 features, we then ran a principal component analysis (PCA) \citep{Pearson_PCA}, which is a dimensionality reduction technique that takes as input $n$ number of variables (in our case the various lightcurve statistics) and performs a linear transformation to identify the most meaningful $n$ dimensional basis in which to express the data \citep{2014PCA}. This PCA is an important step as it helps to better partition the classes in our feature space for more accurate classification. Illustrating the distribution of the first principal component (Figure~\ref{pca_visual}, 500 per class), for example, clearly illustrates how well this feature differentiates between microlensing and other type of lightcurves, in particular those of constants and variable stars. The overlap between CV and microlensing appears more prevalent, but this is expected given how similar these lightcurves appear to one another. 

While Figure~\ref{pca_visual} only illustrates the distribution of the first principal component, visualizing the entire feature space is a more complicated task as we're dealing with a 47 dimensional-space. Stochastic Neighbor Embedding (SNE) is a nonlinear dimensionality reduction technique developed by \citet{hinton2003stochastic} that preserves local data structure and allows for the visualization of high dimensional datasets into a two or three dimensional feature space. SNE works by converting the Euclidean distance of vectors around each point in the original basis to conditional probabilities, such that this probability is high for points near the source-point and lower for points farther away. The mapping is done by modeling the lower-dimensional space in the same manner and minimizing the mismatch in the probability distributions between the two basis. \citet{maaten2008visualizing} built upon SNE and presented a new technique called t-Distributed Stochastic Neighbor Embedding (t-SNE) that instead uses long-tail distributions, thus allowing the lower dimensional representation to be spread more evenly. 

To compare how well our lightcurve classes are partitioned given our original 47 features we simulated 2000 lightcurves (500 of each class) using OGLE cadence as described in Section~\hyperref[hyperparameteroptimization]{5.2}. We simulated a large number of events as we want to create a large feature space to properly visualize how well our features distinguish between our different classes, and while a larger training set better captures the feature space, we investigated the impact of training set size on performance and found performance variance to be negligible when we trained with 500 of each class versus when we trained with several thousand. The 47 metrics described in Table~\ref{features_table} were computed per each lightcurve, after which a PCA was performed to yield 47 principal components. Figure~\ref{tsne_1} displays a 2-dimensional projection of the original feature space composed using t-SNE, which illustrates how classes aren't clustered in an ideal manner given the noticeable overlap between constants and variables, as well as that of ML and CV. The only class that doesn't overlap with microlensing is constant, suggesting that in principle no constants sources should ever be confused by microlensing, and vice versa; though in practice this will depend on the quality of the data and on the magnification of the event.
	
Figure~\ref{tsne_2} illustrates a similar t-SNE projection but using the 47 principal components from the PCA. While the overlap between classes still present as in Figure~\ref{tsne_1}, the classes are overall better clustered with the PCA feature space with a clear distinction existing between ML/CV and constants/variables. Employing a PCA allows us to better distinguish between microlensing and other sources, as unlike Figure~\ref{tsne_1}, the feature space is better clustered according to individual classes. Given that our principal components serve as better features for distinguishing our source classes, applying a PCA transformation to our statistical metrics when classifying new lightcurves will result in higher classification accuracy. While Figure~\ref{tsne_2} still displays overlap between constants and variables, the overlap between CV and microlensing is minimal indicating that our classifier will yield few false-alerts when tasked to distinguish between eruption events and microlensing.
	
While a PCA can yield $n$ maximum components, training a machine learning algorithm with too many components is typically not a good idea as it is important to minimize the amount of noise captured in the feature space, and including more components than necessary can result in overfitting. To identify the ideal number of principal components to utilize, we conducted a hyperparameter optimization as described in the following section.
	
\subsection*{5.2. RF Hyperparameter Tuning}
\label{hyperparameteroptimization}

To output the best possible performance, it's important to optimize the hyperparameters for the task at hand. Each algorithm functions with a set of parameters which ultimately determine classifier accuracy when new, unknown data is input for classification. In the case of the Random Forest technique these parameters are set before training, and include learning parameters such as tree depth and number of features to consider during individual splits, which are in a sense the `settings' at which the classifier operates. These will always require careful tuning to ensure maximum accuracy, as the optimal values may vary depending on the particular dataset. The random forest hyperparemeters we chose to tune are similar to those explored by \citet{Pashchenko2018}, and include:
\begin{itemize}
	\item $n\_estimators$: The number of trees in the ensemble. 
	\item $max\_depth$: The maximum depth allowed per individual tree.
	\item $max\_features$: The maximum number of features to consider when searching for the best split of the node. 
	\item $min\_samples\_split$: The minimum number of data points required to split a node within an individual decision tree.
	\item $min\_samples\_leaf$ : The minimum number of samples required at a leaf node.
\end{itemize}
Determining the best values for these parameters is really an experimental task, as besides the number of trees in the ensemble (more is better), it's impossible to predict just what combination of values in the hyperparameter space will yield maximum accuracy. Therefore we performed a systematic grid search to identify the ideal hyperparameters by tasking the algorithm to classify real microlensing lightcurves, and assessing performance for each combination of hyperparemeters used. As the Optical Gravitational Lensing Experiment (OGLE) is one of the microlensing surveys in longest continual operation, they currently possess the largest catalog of microlensing events. Given the quantity and quality of OGLE's microlensing data and the science that has been derived from it, we recognize the importance of creating a machine learning classifier that can correctly identify OGLE events. The OGLE-II microlensing catalog is, at the time of writing, the largest public catalog of microlensing lightcurves including well over one hundred events, and as such we utilized this sample of events as a means of testing classifier performance \citep{2000Wozniak}. From 162 microlensing events identified in the OGLE-II survey, we visually inspected each event so as to omit lightcurves that appeared to be misclassified either due to poor sampling or noisy data. This left us with 151 lightcurves that we considered credible signals that a classifier should be able to identify as microlensing. As outlined in Section~\hyperref[trainingset]{4}, we used adaptive cadence to create a training set for OGLE-II. This was done by simulating each lightcurve (CV, variable, ML and constant) with randomly selected timestamps from the 151 OGLE-II lightcurves. 
\begin{figure*}
	\includegraphics[width=18.0cm,height=7cm]{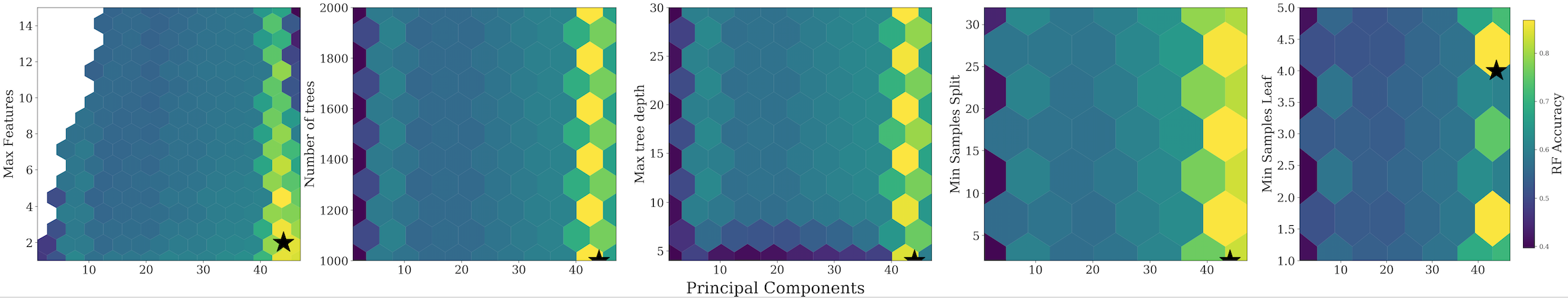}
	\caption{Hyperparameter optimization heatmap. We observe high performance when 40+ principal components are used regardless of what hyperparameter is being optimized. The black stars represent the values for the given parameter that's been integrated into our algorithm. The slant in the first plot is due to the allowable maximum features being limited to the number of principal components being used during a given grid search. 
	}
	\label{pca_heatmap}
\end{figure*}
\begin{figure}
	\includegraphics[width=8.4cm,height=7.0cm]{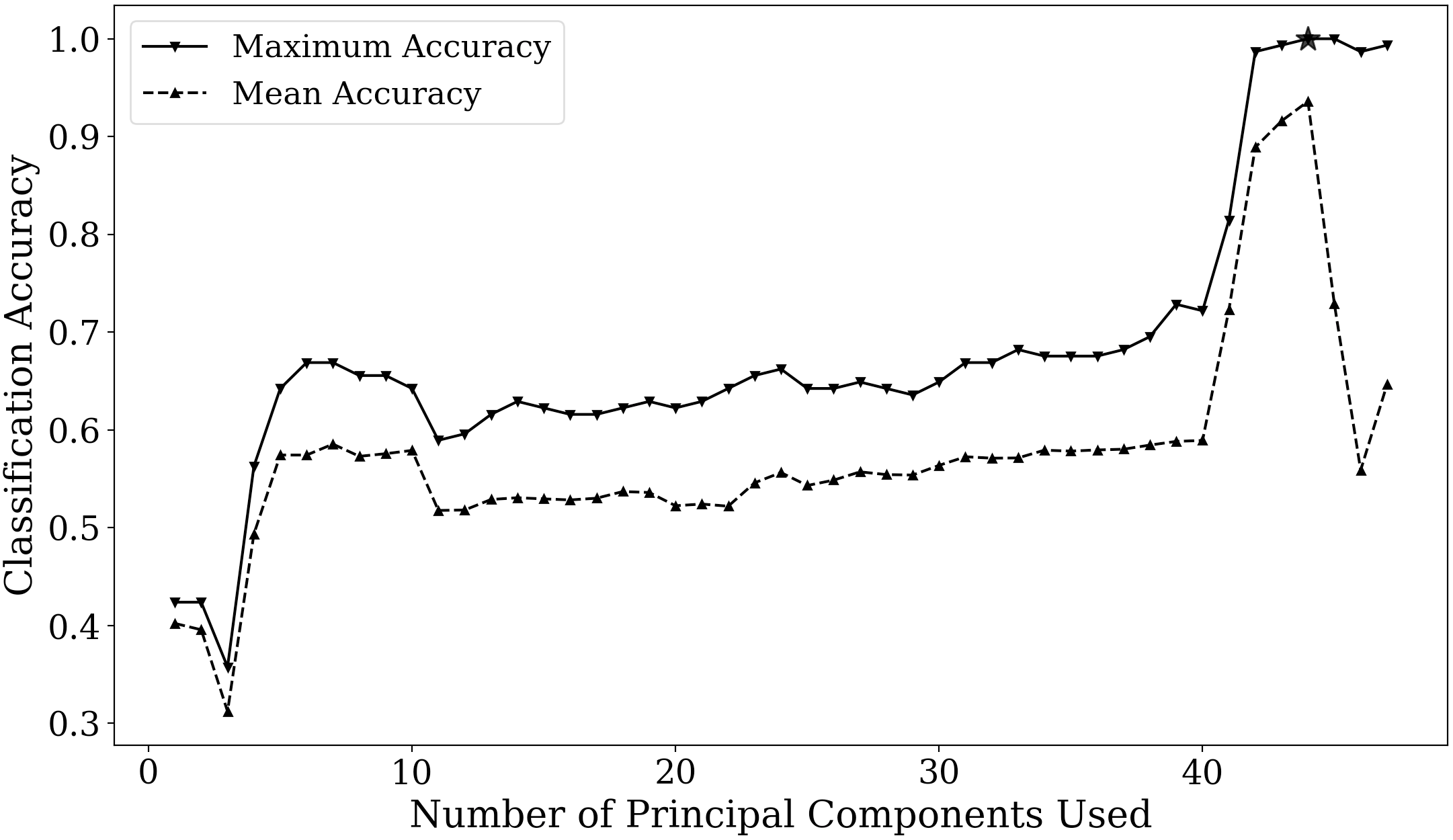}
	\caption{Classifier accuracy as a function of principal components used. The black star is positioned where we achieve maximum classification accuracy, with an OGLE-II detection efficiency of 100\% occurring when 44 components were used. We note that when we make subsequent training sets and train the algorithm with these parameters, this efficiency averages to $\approx$94\%.
	}
	\label{pca_graph}
\end{figure}
We simulated 500 lightcurves per each class, adding noise from a noise-model constructed using PTF data, as described in detail in Section~\hyperref[iptfsearch]{7.2}. Ideally one would simulate a training set with both adaptive cadence and an appropriate noise-model given the particular survey and expected seeing conditions, but constructing an appropriate OGLE-II noise-model would require more information about the telescopes, detectors, observing strategy and systematics than is currently publicly available. As OGLE-II utilized the 1.3 meter Warsaw telescope \citep{1997_OGLEII} and PTF used the 1.2 meter Palomar Samuel Oschin telescope \citep{cuillandre2000performance}, both telescopes should have similar noise properties to first order and thus we applied a PTF noise model to our simulated OGLE training set. Overall we find that PTF data was usually noisier than the subsample of microlensing events extracted from OGLE-II, and in turn we can expect that training the algorithm with lightcurves simulated using OGLE-II cadence but with PTF noise will yield decreased classification accuracy. In this sense, it represents a worse case scenario of how our algorithm performs when tasked to identify OGLE-II events. Despite this limitation, in the following section we demonstrate how we can still achieve high accuracy classifying OGLE-II microlensing lightcurves in this manner -- but nonetheless we emphasize the importance of mimicking survey conditions as accurate as possible for optimal classification results. 

Using the 151 OGLE-II lightcurves as a means of judging classifier accuracy, we configured our random forest hyperparameters by performing the following grid search and recording how many of the 151 events were correctly classified given each combination of parameters.
\begin{itemize}
	\item $n\_estimators$: \{1000 ,2000\} steps=100
	\item $max\_depth$: \{4, 30\} steps=2
	\item $max\_features$: \{1, 15\} steps=1
	\item $min\_samples\_split$: \{2, 32\} steps=5
	\item $min\_samples\_leaf$ \{1, 5\} steps=1
\end{itemize}
This grid search was performed iteratively by initially training and testing the algorithm using only the first principal component, after which the second was added and the grid search began anew, continuing until the maximum $n$ principal components were used. This allows us to not only identify the optimal combination of RF hyperparameters, but also to empirically deduce how many principal components should be used to maximize accuracy. Using 30 cores in a 40x Intel Xeon CPU E5-2630L v4, 1.80GHz processors with 25MB cache, this process took 15.5 days to complete, and while there are other more time-efficient means of optimizing hyperparmeters, such as a random grid search \citep{bergstra2012random}, we chose a standard grid search despite the intensive computation required to increase the likelihood of identifying the ideal combination of hyperparameters that maximized algorithm performance when tasked to classify real microlensing data. 

The heatmap in Figure~\ref{pca_heatmap} demonstrates how performance is most influenced by the number of features one uses for training, with Figure~\ref{pca_graph} illustrating how high accuracy occurs when more than 40 components are used for training, but decreases steadily as it reaches 45 or more. This is expected, as each successive principal component being orthogonal to the rest encapsulates less and less information about the feature space and ultimately begins to capture noise.

From our grid search, we found maximum accuracy when we trained with 44 principal components (marked by the black star in Figure~\ref{pca_graph}), with over 12000 combinations of parameters yielding the same maximum accuracy of 100\%. We then chose the hyperparameters that optimized computation time. Since we can achieve maximum accuracy with 1000 trees, training with more is unnecessary and only increases the time it takes the algorithm to run as more trees must output a prediction before a decision is made. Likewise, while we can achieve maximum accuracy with a tree depth of 30, we can also achieve this with a tree depth of 4; which may also result in quicker classification as votes are output more quickly. From the 12000 combinations, we ultimately chose the following for general application. 
\begin{itemize}
	\item $n\_estimators$: 1000
	\item $max\_depth$: 4
	\item $max\_features$: 2
	\item $min\_samples\_split$: 7
	\item $min\_samples\_leaf$: 4
\end{itemize}
These hyperparameters were encoded into our algorithm and contributed to the performance demonstrated in the following sections.
\subsection*{5.3. OGLE-II}
\label{ogle_ii}
\begin{figure}
		\includegraphics[width=8.4cm,height=8.4cm]{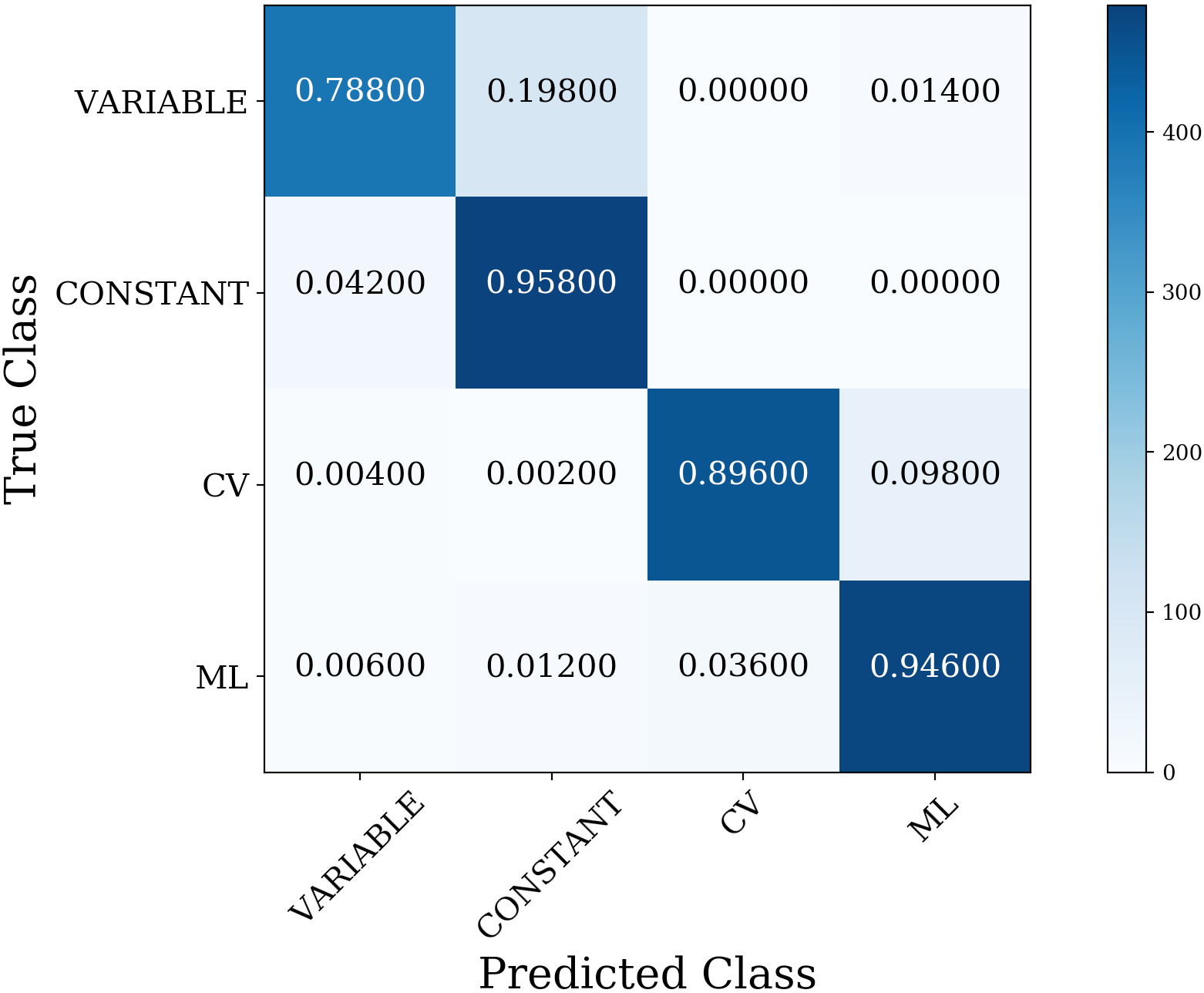}
		\caption{Confusion Matrix. This figure displays algorithm performance when tasked to classify simulated lightcurves.
		}
		\label{confusion_matrix_}
\end{figure}
\begin{figure}
		\includegraphics[width=8.4cm,height=7.0cm]{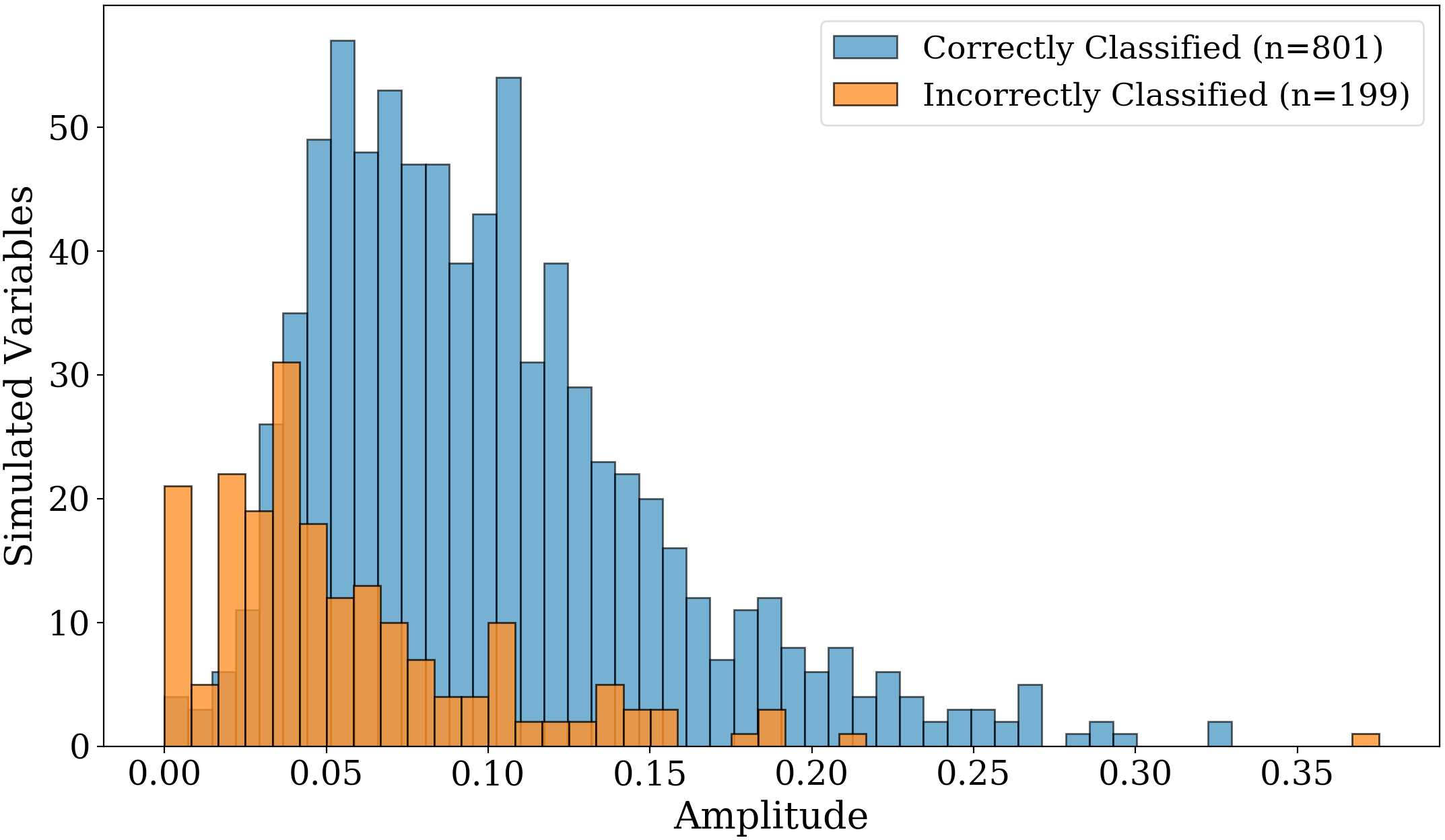}
		\caption{Amplitude distribution of simulated variables. The distribution of variables that are mostly misclassified as constant stars is skewed to the right, as these signals are simulated with smaller amplitude and thus mimic noisy constants.
		}
		\label{amplitude_distribution}
\end{figure}

Training a random forest classifier with these hyperparameters, we then tested our optimized classifier against a simulated dataset produced as described in Section~\hyperref[trainingset]{4}, using OGLE-II cadence and PTF noise. We simulated 500 lightcurves per class, 250 of which were used for training and 250 for testing. Figure~\ref{confusion_matrix_} illustrates classifier accuracy per class, with high overall accuracy although this is expected given that our training set is very much representative of the data we're testing. The only discrepancy is between the constant and variable classes, with a considerable amount of variable lightcurves being misclassified as constant stars. This mismatch occurs when simulated variables appear as noisy constants as a result of being low-amplitude signals, and is consistent with the clustering overlap between these two classes as displayed in Figure~\ref{tsne_2}.

Figure~\ref{amplitude_distribution} portrays the distribution of 1000 simulated variable sources differentiating between those that are classified correctly and those that are classified as constants. Those that are correctly identified as being variable signals tend to have higher amplitude, whereas the ones incorrectly classified have amplitudes less than $\sim$0.05 mag and being low amplitude these sources can display characteristics of noisy constants especially when there's low signal to noise. While this demonstrates the limits of our classifier in detecting low-amplitude variables, this confusion does not impact our search for microlensing signals as Figure~\ref{confusion_matrix_} illustrates minimal confusion ($\sim$1\%) between variables and microlensing.

\section*{6. Early Detection}
\label{earlydetection}
\begin{figure}
	\includegraphics[width=8.4cm,height=7cm]{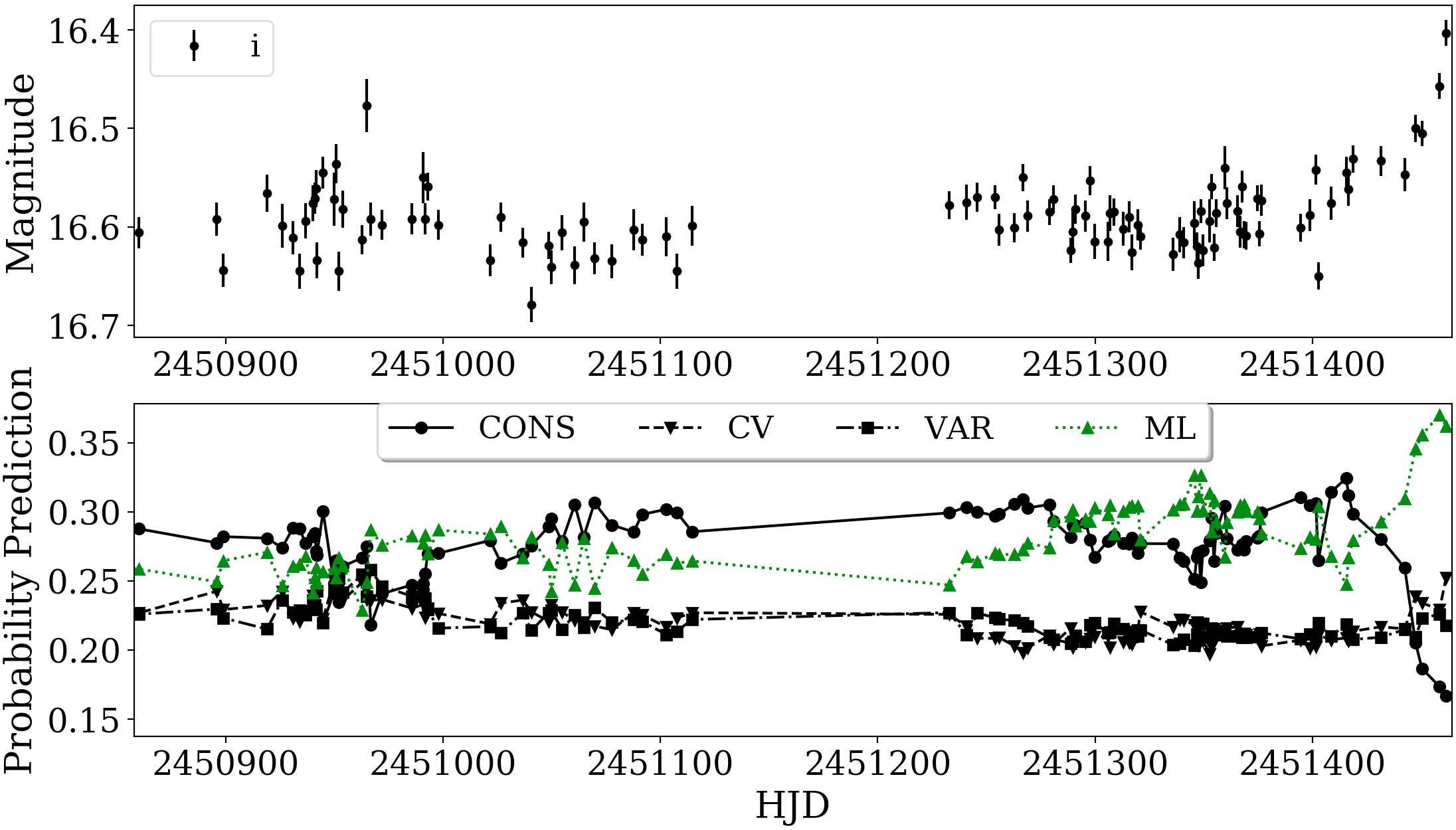}
	\caption{OGLE-BUL-1999-46. Top: OGLE-II microlensing lightcurve displaying characteristics of a single lens event. Bottom: Drip-feeding analysis reveals at which epoch our classifier would have identified this as a microlensing event.  
	}
	\label{ogle_99_46}
\end{figure}
\begin{figure}
	\includegraphics[width=8.4cm,height=7cm]{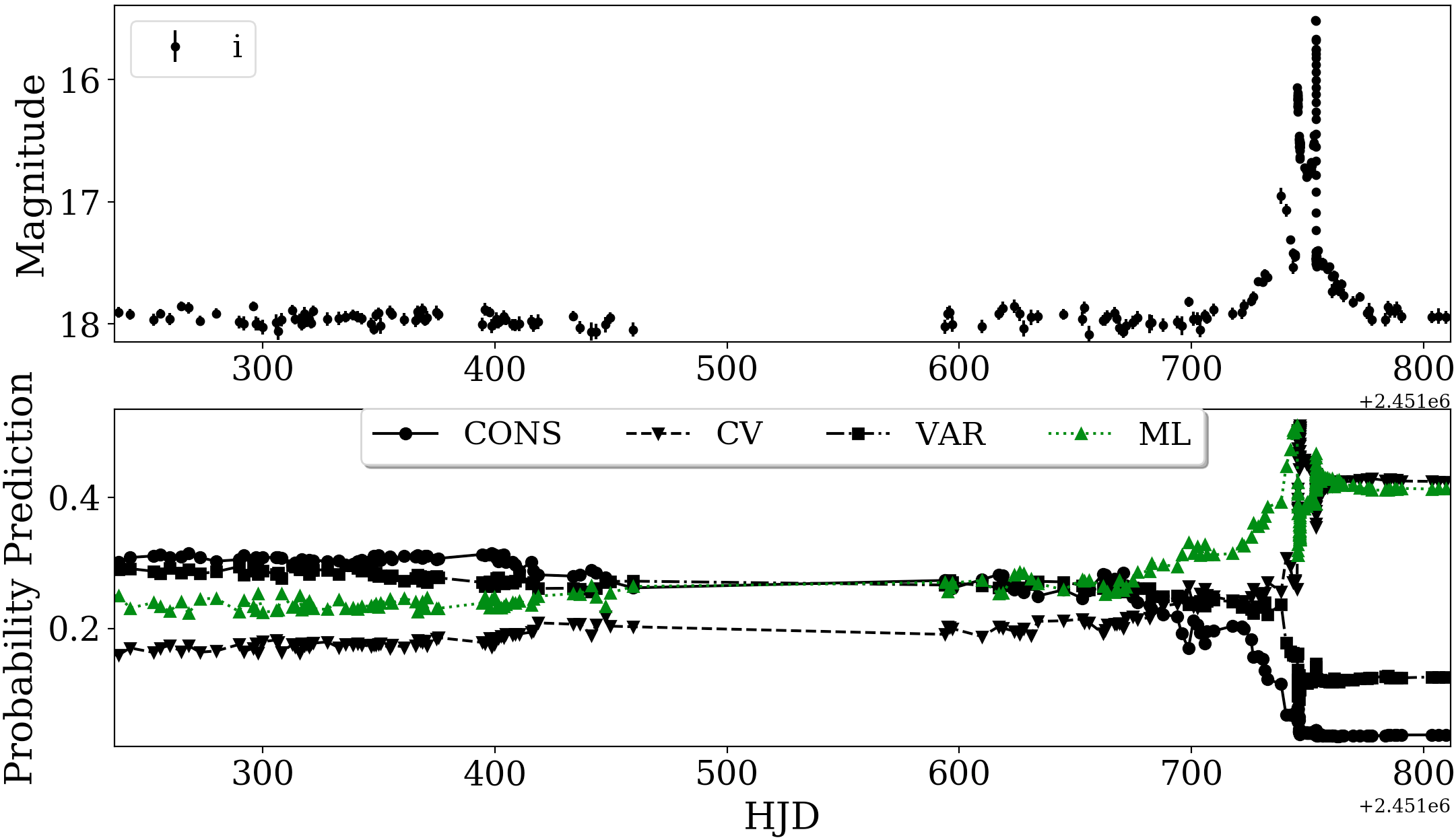}
	\caption{OGLE-BUL-2000-46. Top: OGLE-II microlensing lightcurve displaying characteristics of a binary lens. Bottom: Drip-feeding analysis reveals at which epoch our classifier would have identified this as a microlensing event.  
	}
	\label{ogle_5}
\end{figure}
As rare transient events, microlensing must be followed up as quickly as possible so that enough data can be collected so as to properly constrain the event parameters. As illustrated in Figure~\ref{confusion_matrix_}, we can achieve high accuracy when entire lightcurves are input for classification, but to assess the live detection capabilities of our classifier we devised a drip-feeding process in which lightcurves are classified one point at a time, starting with at least three points, after which the next point is added and the lightcurve is classified again. This drip-feeding analysis allows us to asses algorithm performance in real-time as it reveals the epoch at which microlensing events are classified as such. 

An example of this procedure is demonstrated in Figure~\ref{ogle_99_46}, which displays an OGLE-II single-lens event and how our algorithm flags this event as microlensing during the initial rise of the signal -- we also demonstrate in Figure~\ref{ogle_5} an OGLE-II binary microlensing signal which was among those not correctly classified by our algorithm. Drip-feeding this lightcurve into the algorithm reveals that our classifier is capable of detecting the event right as it begins, indicated by the microlensing probability prediction being higher than all others -- although when the entire lightcurve is input for classification and the binary nature of the source is accounted for, the classifier predicts this to be CV with a probability that's very similar to that of microlensing, indicating high algorithm uncertainty. Confusion in this case is expected, as we have, for the time being, concentrated on single-lens events as incorporating the bold and subtle features that characterize binary microlensing is an ongoing challenge that will be the focus of future research. Nonetheless, Figure~\ref{ogle_5} illustrates how our classifier can handle binary microlensing reasonably well prior to the presence of any anomalies -- a promising performance given our goal of detecting and following-up microlensing behavior as early as possible.

Even though this binary event is not consistently classified as microlensing by our classifier, Figure~\ref{ogle_5} displays how the signal is always classified as microlensing during the rise. Being able to detect these events early on is important, as $\approx$10\% of microlensing events toward the Galactic Bulge can be characterized by binary lens behavior \citep{binary_lens}. Correctly distinguishing between binary and single-lens events will be an important avenue of future development, and even though the current version of this software is tuned for single-lens detection only we nonetheless demonstrate the predictive power of our algorithm during the rising phase of multi-lens events. 

\begin{figure}
	\includegraphics[width=8.4cm,height=7cm]{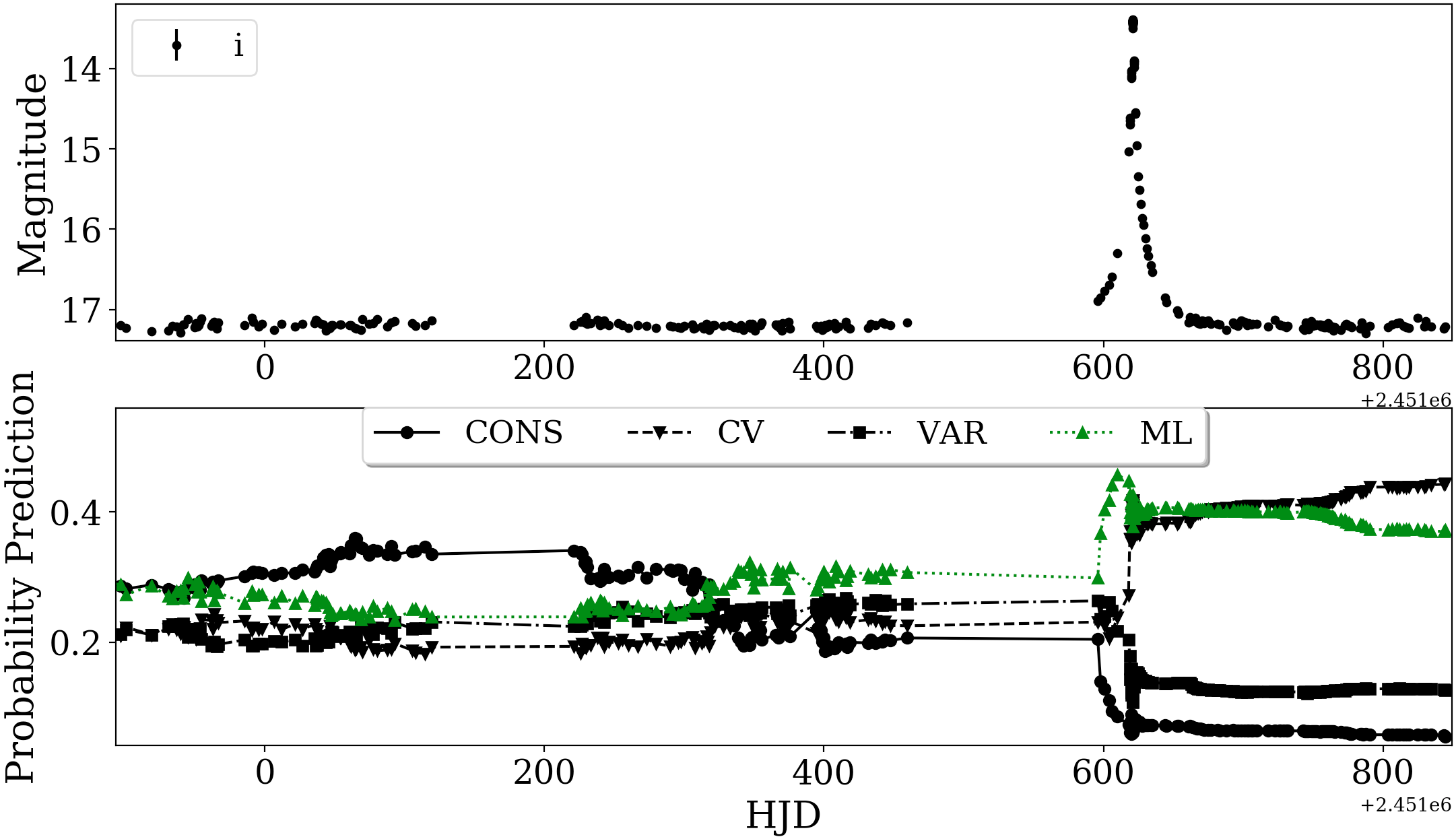}
	\caption{OGLE-BUL-2000-06. Top: OGLE-II microlensing lightcurve incorrectly classified. Bottom: Drip-feeding analysis reveals that this lightcurve was predicted to be ML during the rise and was later misclassified to CV.
	}
	\label{ogle_2}
\end{figure}

\begin{figure}
	\includegraphics[width=8.4cm,height=7cm]{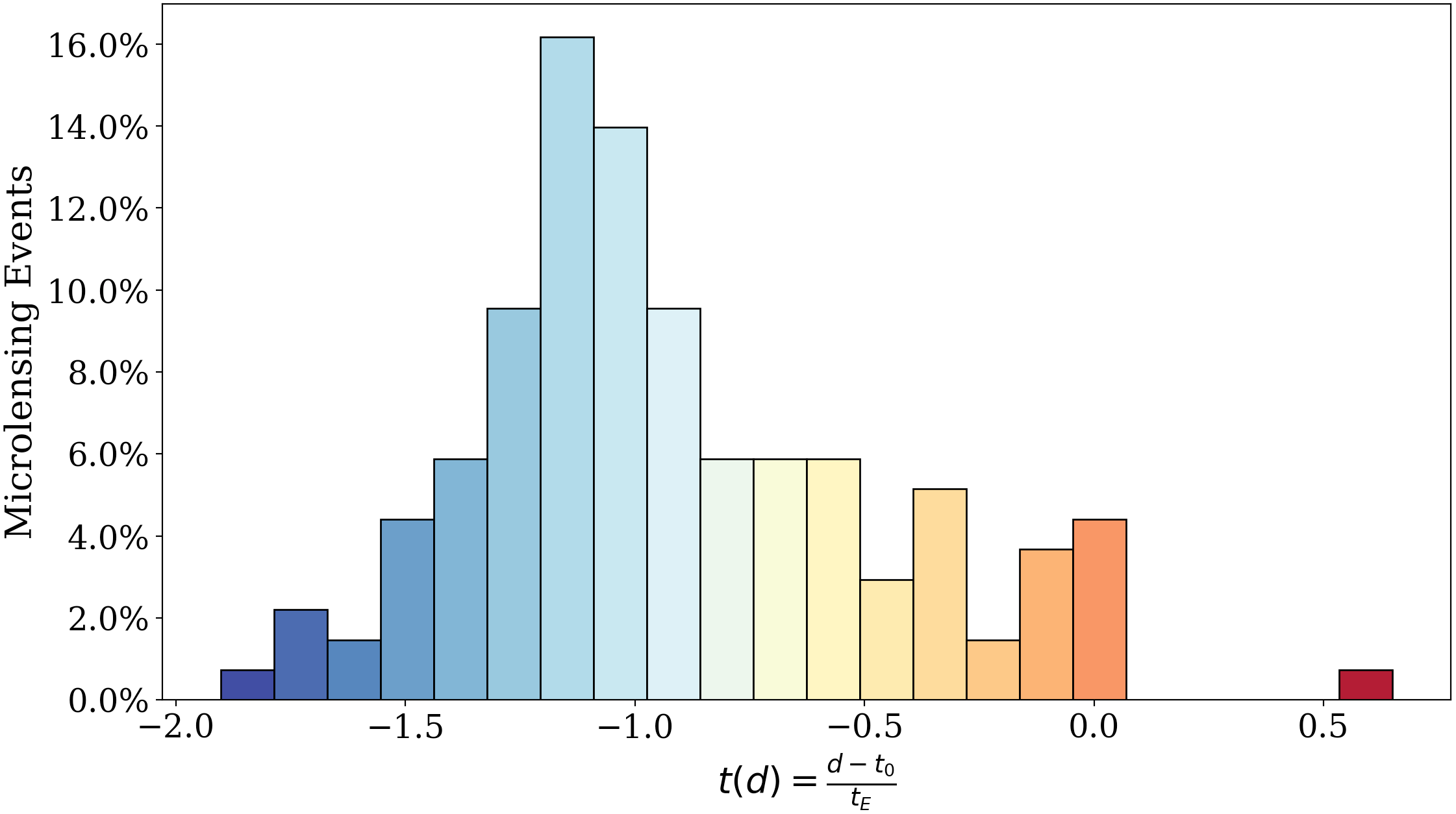}
	\caption{Early Detection Capabilities. Histogram displaying the distribution of $t(d)$ as recorded during the drip-feeding process. We find that our classifier is able to correctly classify most of the OGLE-II microlensing events prior to $t_0$ ($t(d)$ = 0).
	}
	\label{ogle_hist}
\end{figure}

We conducted the same drip-feeding analysis on the $\approx$5\% of OGLE-II microlensing lightcurves that were not classified correctly, and found that all of these events were classified correctly during the rise, and only after the event did the algorithm confuse these for CV. That our algorithm detected all of these lightcurves as microlensing but later misclassified them can be explained by the data-gaps in the photometry, especially during the rising phase of the events (an example lightcurve is displayed in Figure~\ref{ogle_2}). The asymmetry in the lightcurve can confuse the algorithm into mistaking the overall shape of the signal as having the sharper rise and longer tail that commonly characterizes CV outbursts; which is not consistent with the smoother microlensing profile of the single-lens events we simulated and used for training. 
To better illustrate the early-detection capabilities of our classifier, we conducted a drip-feeding analysis of all 151 OGLE-II events, and recorded the following metric per lightcurve during the drip-feeding procedure:
\begin{ceqn}
\begin{align}
t(d) = \frac{d - t_0}{t_E},
\end{align}
\end{ceqn}
where $d$ is the timestamp at which the classifier correctly identifies the event as microlensing. Thus, this metric, $t(d)$, is 0 when the classification occurs at the peak, $t_0$, and $\pm$1 when it occurs at the end or start of the event ($t_0 \pm t_E$), respectively. A high-performing classifier would detect events on or before the rise, thus an ideal value for $t(d)$ would be $\leq-1$. We find this to be mostly true for our classifier as the majority of OGLE-II lightcurves are correctly identified prior to the rise, with only one event identified post-$t_0$ (see Figure~\ref{ogle_hist}). That every misclassified OGLE-II lightcurve is at some point correctly identified as microlensing reveals that we would have correctly detected all OGLE-II microlensing events in real-time had our software been operating live, which also includes instances of binary microlensing events.

\section*{7. Application to Wide-Field Survey Data}
\label{application}

\subsection*{7.1. PTF/iPTF}
\label{ptf_iptf}

The PTF survey began in 2009 as the Palomar Transient Factory (PTF) making use of the wide-field 48-inch Palomar Samuel Oschin telescope to explore variables and transients. The survey utilized the CFH12K camera \citep{cuillandre2000performance}, which provided an 8.1 square-degree field of view with 1'' sampling. Observations occurred in two broad-band filters (Mould-R and SDSS-g'). Given median seeing conditions in Palomar (1.''1) the camera achieved 2'' FWHM images, reaching 5$\sigma$ magnitudes of $m_R \approx20.6$ and $m_{g'} \approx21.3$ in 60 s exposures \citep{Law2009, rau2009exploring}. A real-time image-differencing pipeline issued alerts of transient candidates for follow-up observation (for more information, see \citet{Law2009}).
	
PTF was succeeded by the intermediate Palomar Transient Factory (iPTF) in 2012, making use of the same telescope in Palomar but with better data reduction and classification software (\cite{Kulkarni}; \cite{CaoIPTF}). As with PTF, the survey footprint is not uniformly sampled, such that the time-domain data across fields is inconsistent as imaging cadence was dependent on numerous factors, including visibility at time of year and what sub-survey each field belonged to \citep{price2014statistical}. Finding microlensing in low-cadence fields is challenging as the events occur only once, and gaps in data can make definite classification impossible. While we do not expect to fully characterize microlensing events using iPTF data alone, we seek to demonstrate that even in an irregularly sampled survey, we can still detect microlensing early enough that additional data can be obtained with follow-up telescopes. 

\subsection*{7.2. iPTF Search}
\label{iptfsearch}
As the PTF/iPTF data is currently undergoing reprocessing efforts, we did not have access to all of the data, which included data from the most promising regions of the sky such as the Bulge and Galactic Plane. We searched the currently available data, training a machine learning classifier with PTF cadence. We utilized a subset of $\approx$10000 PTF lightcurves made public by \citet{price2014statistical} to create a PTF noise-model. This was done by fitting a third degree spline curve to the $log_{10}(\text{RMS})$ as a function of median magnitude for the released lightcurves, with our saved fit allowing us to add an appropriate amount of noise to a lightcurve given its baseline magnitude. 
\begin{figure}
	\includegraphics[width=8.4cm,height=7cm]{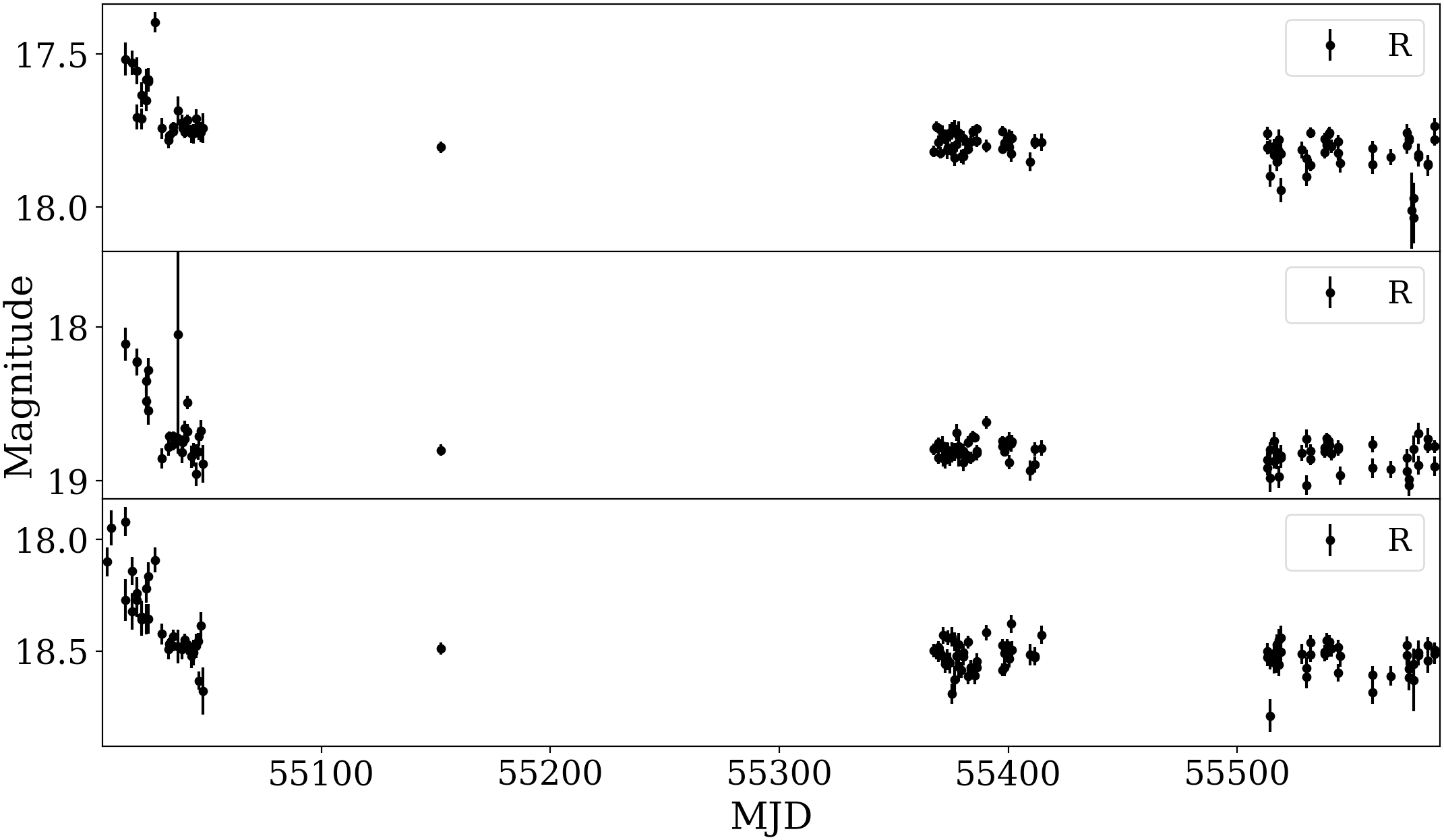}
	\caption{Example lightcurves of microlensing candidates detected in iPTF data. Given that the increase in flux occurred at the same time for these lightcurves, in conjunction with the full moon present during this epoch suggests that these are unlikely to be true microlensing signals.    
	}
	\label{iptf_1}
\end{figure}
We simulated PTF lightcurves with baseline magnitudes ranging from 14 to 20, producing 500 examples of each of our respective classes. The timestamps at which we simulated each lightcurve were extracted randomly from the subset of $\approx$10000 PTF lightcurves, allowing us to replicate PTF cadence. We again conducted a PCA to the statistical metrics derived from our simulated PTF lightcurves, and with the PCA transformation saved and the principal components employed to train a Random Forest model, we then input available PTF/iPTF lightcurves for classification. Given the limited availability of iPTF data at the time of research as well as the irregularity of the sampling, we chose to not limit our query based on coordinates and instead we queried all available data regardless of its location in the sky. As such, we queried iPTF data by requiring that lighturves contain at least 45 points as well as a baseline magnitude between 14 and 20, and furthermore we required that no more than 5\% of lightcurve measurements be flagged as bad points as assigned by the iPTF pipeline. If a lightcurve passed these criteria and was input for classification, we imposed an additional probability prediction threshold of at least 0.6 so as to limit the misclassification of lightcurves due to either poor photometry or inadequate data, a threshold previously applied by \citet{Mislis} to filter out poor predictions. In total we inspected approximately 20 million lightcurves, but ultimately only 1.55 million were input into our algorithm for classification as per the imposed photometry and quality conditions. From this subset, approximately 12400 lightcurves were flagged as plausible microlensing signals. To further limit the number of candidates requiring visual inspection for confirmation, we applied pyLIMA, an open-source microlensing modeling package for fitting models to microlensing lightcurves \citep{ebachelet_2017_997468}. We fit each of the selected lightcurves with a PSPL model, and saved only those that fell within the following parameter space:
\begin{itemize}
	\item $10 \leq t_E \leq 50$
	\item $0 < u_0 \leq 1$
	\item Reduced $\chi^2$ < 10
	\item At least two measurements within $t_0 \pm t_E$.
\end{itemize}
We chose to limit $t_E$ to 50 days as a means of avoiding bad fits that stretched across several observing seasons -- imposing all of these conditions to the microlensing model ensures that we only inspect lightcurves that are at least somewhat representative of microlensing behavior, as any variables and eruption events misclassified as microlensing by our algorithm should in principle yield a large $\chi^2$ when fitted with a PSPL model assuming adequate baseline. While pyLIMA alone can be utilized to search for these rare transient events by fitting lightcurves and saving only those that yielded reasonable parameters, it would be computationally inefficient when searching in wide-field surveys that observe tens of millions of stars, and less efficient for sparsely sampled lightcurves where PSPL fits are not well constrained. We report great success when applying pyLIMA as an additional filter, as given the conditions we imposed on the PSPL parameters, we were able to truncate our final sample of microlensing candidates to 200 -- we then visually inspected each of these, selecting 19 total lightcurves that we believed could be true microlensing.

Visual inspection revealed that all lightcurves displayed an increase in flux at approximately the same timestamps (see Figure~\ref{iptf_1} for example lightcurves), and upon closer inspection we found that these increases were occurring during a full moon phase and fitting these lightcurves with a PSPL microlensing model yielded timescales of approximately a month, reflecting the lunar cycle -- this suggests that these lightcurves are the result of systematic noise. Nonetheless, the shape of these lightcurves could indeed represent true microlensing behavior, and as such we expect our algorithm to be able to detect signals such as these which display an increase in flux followed by a relatively flat baseline. Ultimately, we seek to query the most crowded iPTF fields when the data becomes available.
	
\subsection*{7.3. Price-Whelan Events}
\label{price_events}
We compared the performance of our classifier against the analysis conducted by \citet{price2014statistical}, in which he performed a systematic search for microlensing events in PTF data, filtering for true microlensing signals by imposing thresholds on numerous lightcurve statistics. 
\begin{figure}
	\includegraphics[width=8.4cm,height=7cm]{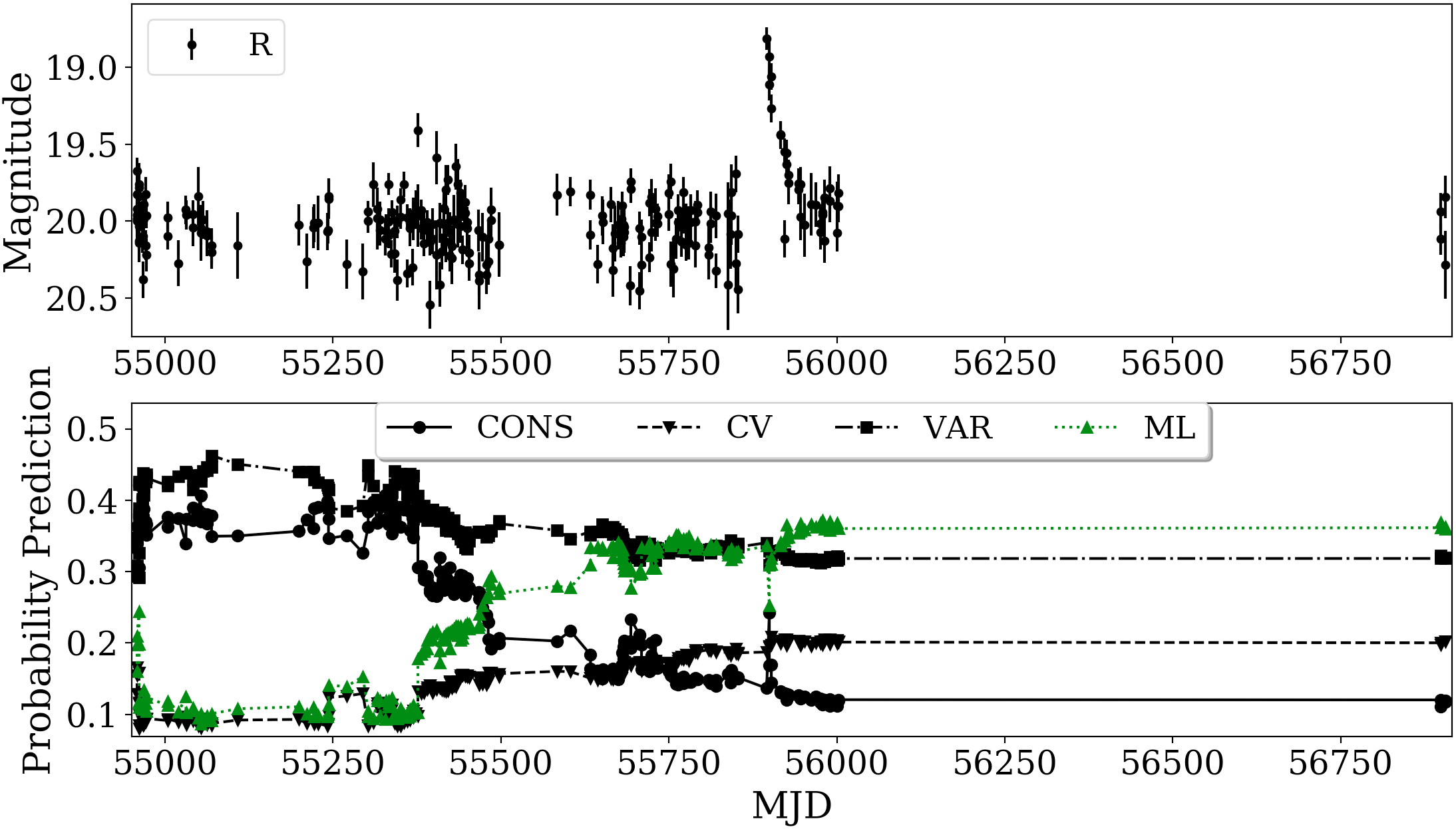}
	\caption{Microlensing candidate PTF1J61502.39. Top: Microlensing candidate detected by \citet{price2014statistical}. Bottom: Drip-feeding the lightcurve into our classifier reveals at which epoch the classifier would have detected the event. 
	}
	\label{price_1}
\end{figure}
We performed a close analysis of the three plausible microlensing events reported by \citet{price2014statistical} to further test the early-detection capabilities of our algorithm. We drip-fed the first candidate into our algorithm (Figure~\ref{price_1}), which illustrates how our classifier would have flagged this source as plausible microlensing right before the peak of the signal, in time to conduct follow-up observations.
\begin{figure}
	\includegraphics[width=8.4cm,height=7cm]{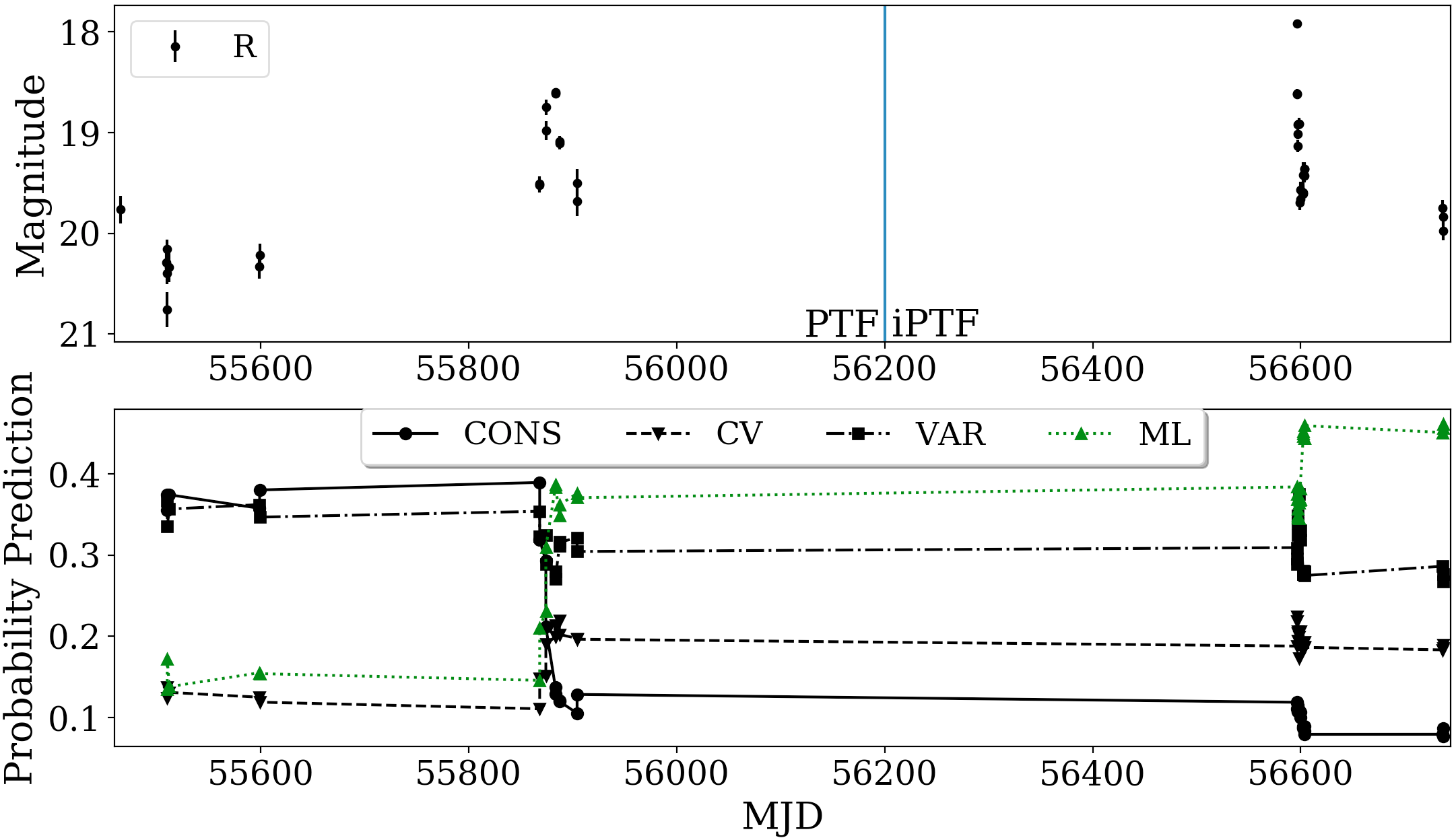}
	\caption{PTF1J061800.25. Top: Microlensing candidate detected by \citet{price2014statistical}. Bottom: Drip-feeding the lightcurve into our classifier reveals at which epoch the classifier would have detected the event. This lightcurve was ruled out as a microlensing candidate when more data became available from the iPTF survey.
	}
	\label{price_3}
\end{figure}
The event PTF1J061800.25 (Figure~\ref{price_3}) was ruled out after iPTF gathered additional data displaying characteristics of a more periodic or outburst-like signal; nonetheless our algorithm flags this as plausible microlensing, illustrating a limitation of our classifier when tasked to classify lightcurve types that are not included in the training set. While it's better to properly account for as many source classes as possible, this is an involved process that ultimately requires vast amounts of good quality data, further complicated given our current inability to account for noisy data and/or mysterious anomalies that may occur. As we trained our algorithm to distinguish between only four classes of lightcurves, simulated with an adequate signal to noise, we can expect confusion and false-alerts when we input noisy data or signals that are not accounted for in our training set, as is the case for PTF1J061800.25.

\begin{figure}
	\includegraphics[width=8.4cm,height=7cm]{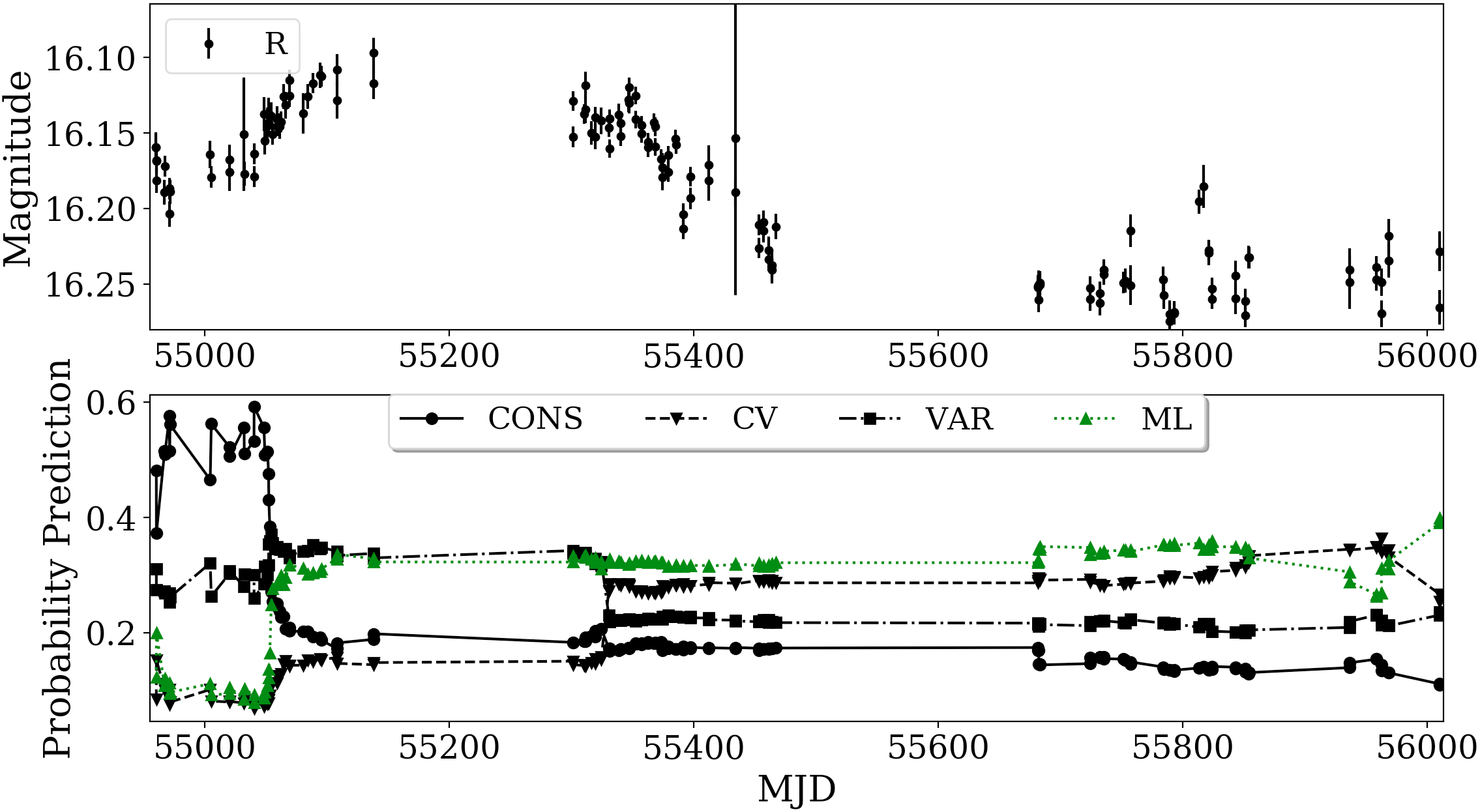}
	\caption{PTF1J172826.08. Top: Microlensing candidate detected by \citet{price2014statistical}. Bottom: Drip-feeding the lightcurve into our classifier reveals at which epoch the classifier would have detected the event. 
	}
	\label{price_2}
\end{figure}

Lastly, candidate PTF1J172826.08 (Figure~\ref{price_2}) remains as plausible microlensing, but as with the other two candidates the data is insufficient to properly determine the nature of this source. We can see how our algorithm flags this event as a variable star during the rise, with microlensing being output as the prediction only during the fall, with the probability of the source being a CV increasing when the lightcurve appears to rise again, but as it returns back to baseline the classifier outputs a final ML prediction for the entire lightcurve. We note that the ML probability prediction for all three of these events is relatively low, hovering over 40\%, so even though the classifier identifies these as microlensing had they been in the subset of iPTF data we queried, these lightcurves would not have passed our 60\% probability threshold and would have been rejected. This suggest that perhaps it's best to be lenient when setting a minimum probability threshold, although it's certainly plausible that these events are not true microlensing with the low ML probability serving as an indicator -- thus imposing a probability threshold may be preferred, although it's difficult to empirically determine the ideal threshold to set without a clear picture of the data quality one can expect, as noisier data may yield lower predictions thus requiring the threshold to be appropriately set for the database at hand.

\section*{8. ZTF}
\label{ztf}
ZTF is a new optical wide-field survey which follows on the success of the PTF/iPTF programs, making use of a similar infrastructure that operated the previous surveys including the same 1.2 meter Samuel Oschin Telescope, but with a new 47 $\text{deg}^2$ survey camera, providing a field of view 6.5 times larger than that of its predecessors \citep{BellmZTF}. Similar to PTF, the ZTF camera achieves image quality of 2'' FWHM, reaching $5\sigma$ magnitude in Mould-R of 20.4 using a 30 s exposure time \citep{BellmZTF}. The pipeline in charge of the astrometric and photometric calibration is very similar to the pipeline that processed PTF data as referenced in Section~\hyperref[application]{7}. The ZTF public Northern Sky Survey began on March 17, 2018, covering the entire visible sky in both $g$ and $r$ every three nights, with the Galactic Plane covered with a one night cadence \citep{2018_ztf_tde}. ZTF data is processed by IPAC where alerts for transients and variables are generated from ZTF images and photometric information (such as a real-bogus score) is provided (for an overview of the ZTF Science Data System, see \citet{ZSDS_Masci}). 
	
While the ANTARES alert broker \citep{Gautham_Antares} is expected to come on-line by the end of 2018 (ANTARES\footnote{https://antares.noao.edu} has come online, update (Dec 2018)), as of time of writing the ZTF alerts are being released every morning as compressed tar archives, containing the full 5-sigma alert stream. Several brokers are currently presenting alerts from the ZTF stream, including the UK-based Lasair\footnote{https://lasair.roe.ac.uk} and the Las Cumbres Observatory MARS\footnote{https://mars.lco.global} interface. The photometric information encapsulated in each alert is available on MARS, such as the star-galaxy classification score assigned using SExtractor during the data processing stage \citep{ZSDS_Masci}, in addition to derived features intended to simplify the filtering process. We make use of this photometric information available in MARS to perform an initial filtering of the ZTF alerts, saving alerts that meet the following criteria:
\begin{itemize}
	\item The latest photometric measurement must be brighter than 17 mag to facilitate follow-up observations. 
	\item The source must get brighter than 0.2 mag from the previous measurement. 
	\item The alert must have occurred within the past seven days.
	\item The lightcurve must have at least five points. 
	\item The real-bogus score (0 is bogus, 1 is real) must be greater than 0.8
	\item The star-galaxy score (0 is galaxy, 1 is star) must be greater than 0.8.
\end{itemize}
In general, searching for microlensing signals at the start of any survey is difficult, with the discovery of new variables and transients being especially problematic as we are unable to utilize sufficient baseline data to omit these false-alerts. While we acknowledge the difficulty in differentiating true microlensing signals from variable sources which lack an adequate baseline, we nonetheless employed our classifier to query ZTF data in anticipation of engaging in follow-up once the survey has sufficient baseline to reliability classify alerts. As with our previous search in PTF data, we created a ZTF training set using adaptive cadence, and given that ZTF achieves similar image quality as PTF, we resort to simulating our classes by applying the PTF noise model described in Section~\hyperref[iptfsearch]{7.2}. The timestamps at which we simulated our ZTF classes were extracted from a simulated ZTF scheduler. We simulated 500 of each source class, and trained the Random Forest classifier using the optimized hyperparameters derived in Section~\hyperref[hyperparameteroptimization]{5.2}.
	
On average around 200,000 ZTF alerts are released each night, with <3000 passing our initial filtering. We then employ an automated search that cross matches the coordinates of these alerts with the Gaia DR2 catalog \citep{GAIA_DR2}, further filtering out alerts which are flagged as variable sources in the Gaia DR2 data release. This approach reduces our initial alerts by approximately one-third, leaving us to insert <2000 alerts into our classifier each day. As the probability that variable stars will undergo microlensing is the same as for any other star, limiting our search to non-variable sources ultimately hurts our chances of detecting true microlensing signals -- but as ZTF is a new survey and no baseline data is available, we are resorting to applying this filter for the first year of ZTF data as data from variables with no baseline can yield a large number of false-alerts. While ZTF provides $g$ and $r$ data, it's not always uniform with some lightcurves containing more data in one band than in another. For this reason every time we input alerts for classification we currently classify the single-band data in which the majority of the photometry is measured in. Through this process our classifier flags <1 \% as plausible candidates that are forwarded along for visual inspection. Microlensing candidates that are deemed credible signals during the visual inspection phase are saved after which we attempt to fit a PSPL model to the lightcurve using pyLIMA to assess its credibility as plausible microlensing.
	
To minimize the risk of false-alerts, a common strategy is to require constant baseline at least before the signal, a technique applied in OGLE-III to minimize the number of false-alerts \citep{2015_Wyrzykowski}. A similar strategy was employed during the beginning of the OGLE-IV phase, as the team chose not to issue alerts for about a year on their website\footnote{http://ogle.astrouw.edu.pl/} on the grounds that the false-alert rate would have been excessive without an adequate baseline. While lack of adequate baseline at the early stages of ZTF is an obvious challenge, we nonetheless demonstrate how our algorithm is capable of detecting plausible microlensing events in real-time.
	
\subsection*{8.1. ZTF18aaveloe}
\label{ZTF18aaveloe}
The microlensing candidate ZTF18aaveloe was first flagged by our classifier on August 2, 2018. Located in a crowded region near the disk of the galaxy (equatorial J2000 RA and DEC in decimal degrees: \{290.5887292, 29.0411098\}), it had no variable tag in any catalogs and very much represented lensing behavior when detected. When we fitted the event with a PSPL model using pyLIMA, the output parameters were very unconstrained, as is common for an event in progress, with a PSPL fit yielding the  parameters in Table~\ref{params1}.
\begin{figure}
		\includegraphics[width=8.4cm,height=7cm]{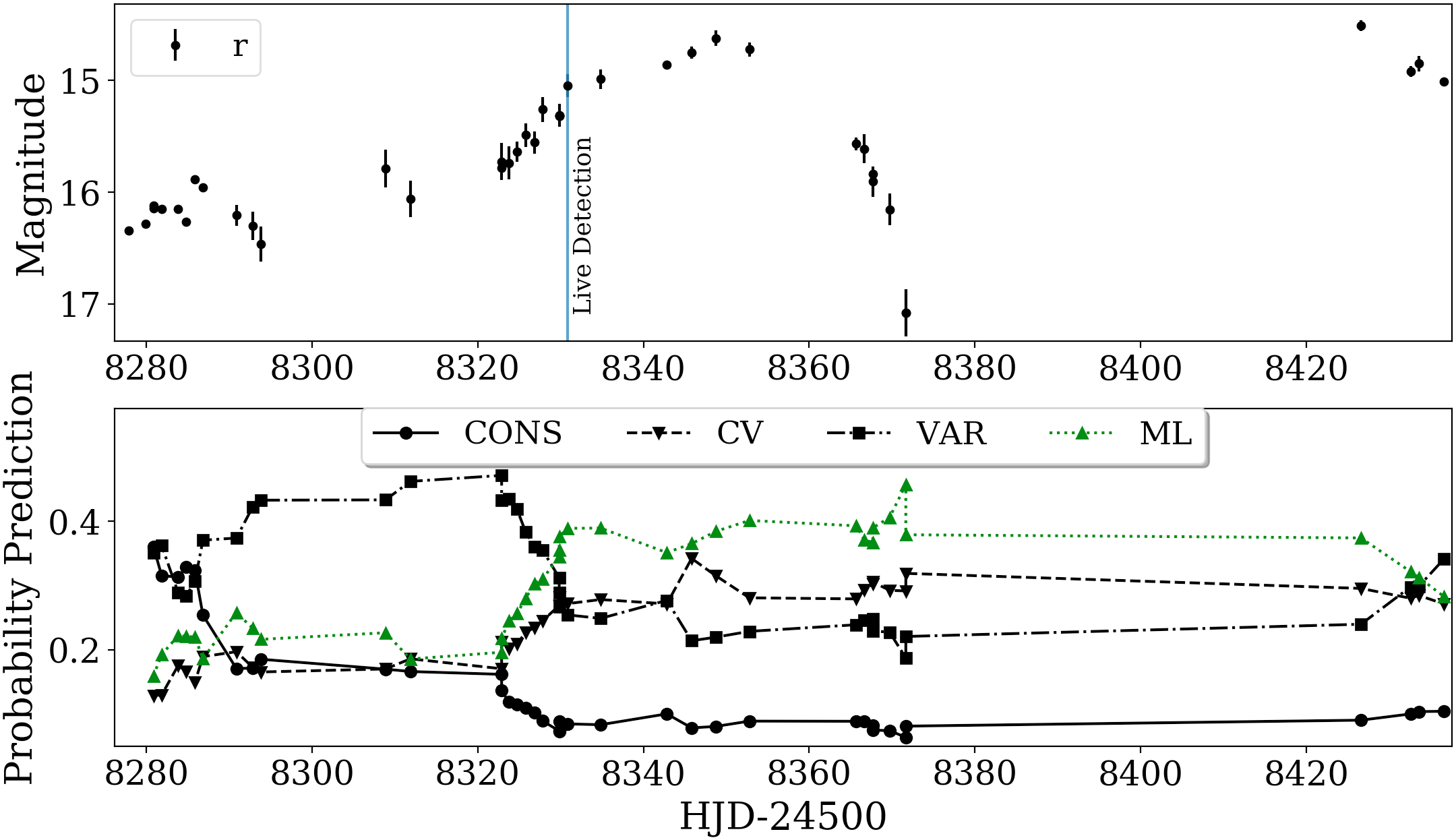}
		\caption{ZTF18aaveloe. Drip-feed of microlensing candidate in ZTF, with the vertical line representing the timestamp at which we detected this in real-time.
		}
		\label{drip1}
\end{figure}
The lightcurve appears to display variability that's not consistent with lensing behavior just prior to the rise, behavior recognized by our algorithm as during this time the lightcurve is classified as a variable star (Figure~\ref{drip1}). We can see that as the source begins to significantly brighten the algorithm prediction progresses toward microlensing, but the probability is relative low at $\approx$35\% throughout the duration of the signal. The last measured data, reverting back to the observed brightening, indicates that this is not true microlensing and rather a variable source, and as expected the microlensing probability prediction quickly drops and a final variable prediction is the output.
\begin{table}
		\begin{center}
			\begin{tabular}{c|c|c} 
				\textbf{Parameter} & \textbf{Fit} & \textbf{Error}\\
				\hline
				$t_0 \ (\text{HJD-24500})$ & 8336.708 & 128.23\\
				$u_0$ & 7.86 $\times 10^{-5}$ & 1296.94\\
				$t_E \ (\text{Days})$ & 62.385 & 1193.93 \\
				g & 5.454 & 184.57\\
				\hline
				${\chi}^2$ & 755.66 \\
			\end{tabular}
		   
			\captionsetup{justification=centering}{\mbox{\textbf{Table 2.} ZTF18aveloe PSPL fit output during live detection.}}
			\captionlistentry{}
            \label{params1}
		
		\end{center}
\end{table}
	
\subsection*{8.2. ZTF18aayczxl}
\label{ZTF18aayczxl}
\begin{figure}
		\includegraphics[width=8.4cm,height=7cm]{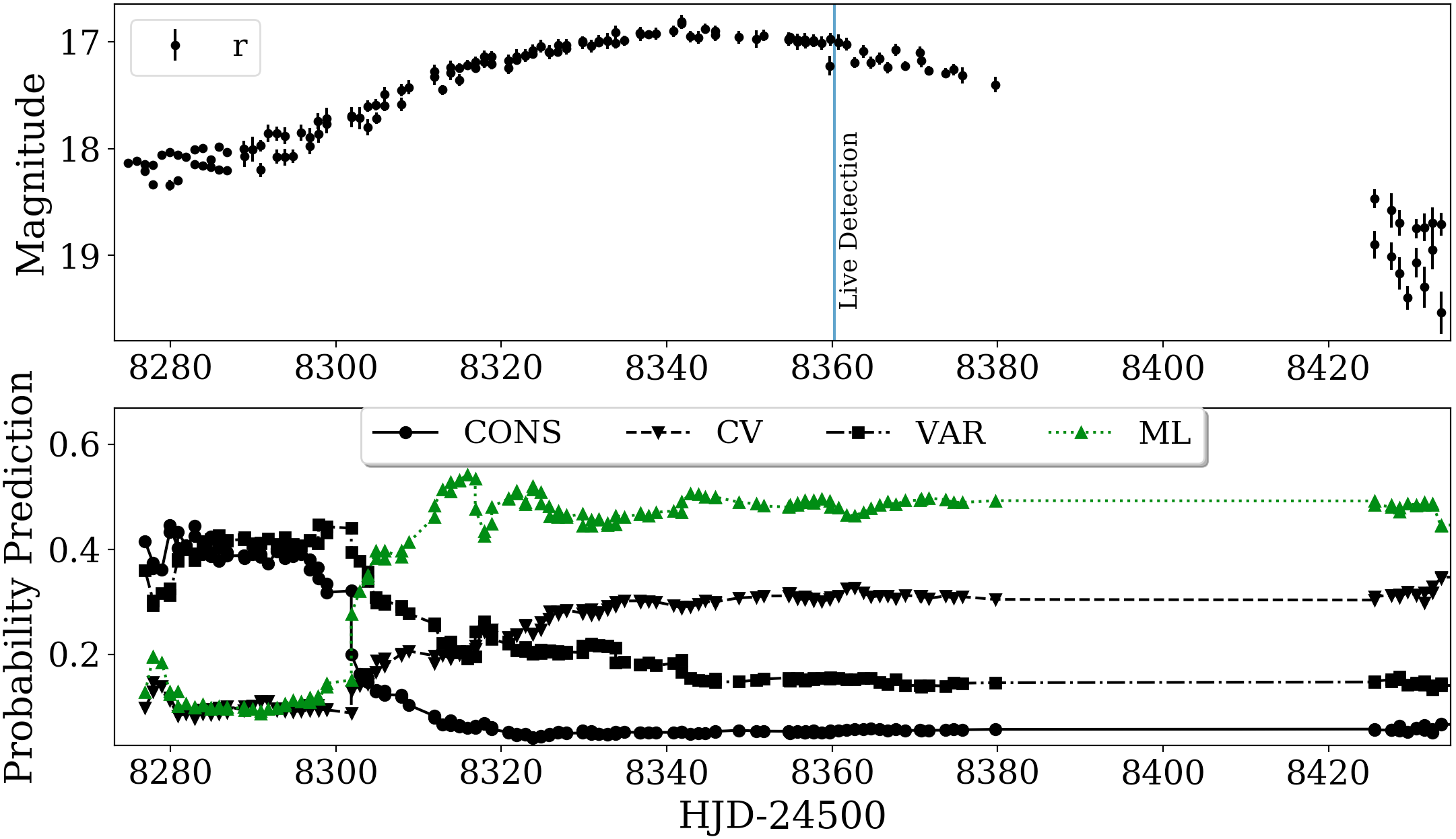}
		\caption{ZTF18aayczxl. Drip-feed of microlensing candidate detected in ZTF, with the vertical line representing the timestamp at which we actually detected this in real-time.
		}
		\label{drip2}
\end{figure}
The source ZTF18aayczxl was identified as a microlensing candidate on August 29, 2018. While the lightcurve is well sampled, the data is quite noisy especially during the rise. We searched the source coordinates (J2000 RA and DEC in decimal degrees: \{292.6685872, 16.0195724 \}) in other catalogs and found no variable tags, with SIMBAD listing this source as a star. Using photometry alone we cannot confidently determine the true nature of this source, but as it truly mimics a microlensing event this is nonetheless the type of lightcurves we sought to detect using machine learning techniques. Even though we detected this event after the peak as per our filtering process, our algorithm classified this signal as microlensing early on in the rise as shown in Figure~\ref{drip2} which displays the drip-feeding process on this lightcurve.
\begin{table}
		\begin{center}
			\label{tab:params2}
			\begin{tabular}{c|c|c} 
				\textbf{Parameter} & \textbf{Fit} & \textbf{Error}\\
				\hline
				$t_0 \ (\text{HJD-24500})$ & 8342.75& 0.847\\
				$u_0$ & 6.075 & 815.38\\
				$t_E \ (\text{Days})$ & 7.78 & 996.15\\
				g & -0.999 & 0.162\\
				\hline
				${\chi}^2$ & 333.07 \\
			\end{tabular}
			
			\captionsetup{justification=centering}{\mbox{\textbf{Table 3.} ZTF18aayczxl PSPL fit output during live detection.}}
			\captionlistentry{}
            \label{params2}
			
		\end{center}
\end{table}
The drip-feeding analysis reveals how our classifier confidently detected the event early on, but due to the unconstrained fit output by pyLIMA (Table~\ref{params2}) and a blending coefficient $g\approx-1$, we suspect this might also not be a true microlensing signal, and as more data is gathered and a proper baseline can be established we seek to make a more definite determination regarding the true nature of this source.

\subsection*{8.3. ZTF18abegegr}
\label{ZTF18abegegr}
\begin{figure}
		\includegraphics[width=8.4cm,height=7cm]{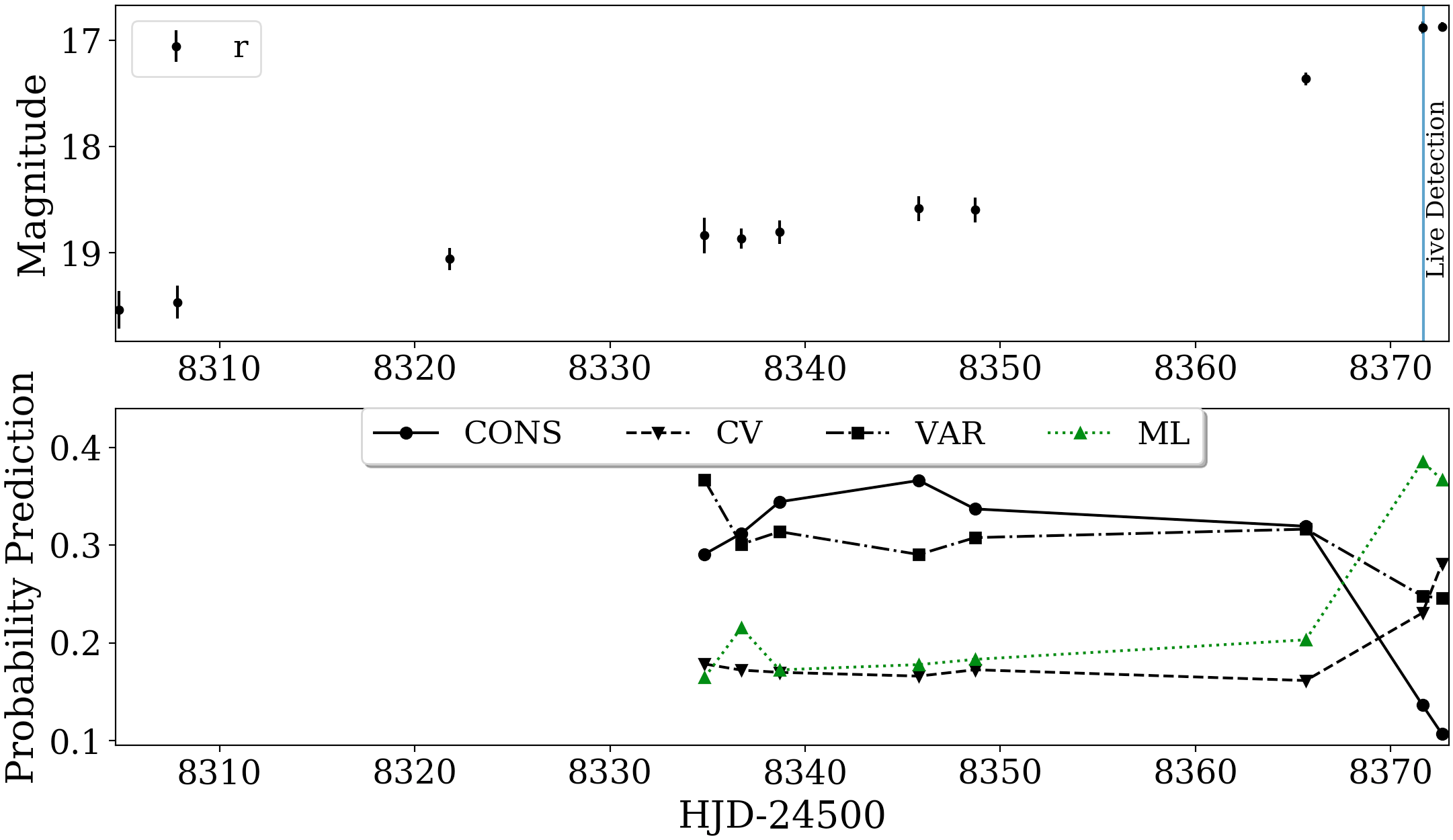}
		\caption{ZTF18abegegr. Drip-feed of microlensing candidate detected in ZTF, with the vertical line representing the timestamp at which we detected this in real-time. We confirmed via the Gaia catalog that this is a variable source. 
		}
		\label{drip3}
\end{figure}
The source ZTF18abegegr was identified as a microlensing candidate on September 10, 2018, with coordinates (J2000 RA and DEC in decimal degrees: \{276.877171, -6.6488872\}). While the available data portrays this to be plausible microlensing, we note that this source is flagged as a variable star in the Gaia catalog, demonstrating the challenge of identifying true alerts when no baseline is available. This lightcurve is representative of true microlensing behavior given its shape and gradual flux increase, and as such is the type of signal we expect our classifier to be able to identify as microlensing.
	
\section*{9. Conclusion}
\label{conclusion}
We have developed a detection algorithm capable of detecting microlensing events in progress from sparsely-sampled lightcurve data. This classifier has been trained using simulated data, tested with existing public data, and successfully applied to the ongoing ZTF public alerts stream. This detection algorithm is an open-source program that has been optimized for microlensing detection in any wide-field survey, with the training set easily modifiable to match the cadence of any given survey (instructions available in the documentation \citep{daniel_godines_2019_2541465}).

We seek to explore avenues of future development that can improve algorithm performance, including implementing code to handle lightcurves that are measured in multiple filters, especially relevant when we begin to search for microlensing in the LSST multi-band footprint. Making use of multi-band photometry would allow us to take advantage of the achromatic nature of microlensing and further filter out false-alerts. Furthermore, we plan to explore the variety of features that can be derived from binary microlensing so as to deduce what metrics would be useful in distinguishing these types of lightcurves. In addition, improving the performance of our classifier by incorporating new, suitable metrics to aid in differentiating between microlensing and other sources will ultimately make our algorithm more robust to problematic cases, such as those presented in Figures~\ref{ogle_5} and~\ref{ogle_2}.

With the ANTARES touchstone now live and operational, we hope to implement our classifier as a new stream into the ANATARES system where its output would become publicly available for review by anyone engaged in microlensing efforts. The code has been designed for easy application across any wide-field survey, and even though it hasn't been tested or optimized for more traditional microlensing survey data, it could still be applied to search for microlensing in public datasets such as UKIRT \citep{shvartzvald2017ukirt} and, as demonstrated, ZTF. Ultimately we hope our efforts in the field of microlensing will lead to the discovery of exoplanets, a promising avenue given the sensitivity of microlensing to small planets beyond the snow line \citep{tsapras2018microlensing}, although such discoveries will require our algorithm to detect microlensing events early enough to trigger follow-up observations as planetary anomalies are very short-lived -- a feat our algorithm is capable of as per the analysis of the classifier performance provided in this study. We will continue to query ZTF data in search of microlensing, with the overall goal of maximizing classifier performance by the time the LSST era begins. 
	
\section*{10. Acknowledgements}
\label{acknowledgements}
The authors gratefully acknowledge funding from LSSTC's Enabling Science program, which enabled work by DG, and from NASA grant NNX15AC97G, which supported work by EB and RAS. This work made use of the OGLE-II and PTF/iPTF photometry catalogs.
	
Based on observations obtained with the Samuel Oschin Telescope 48-inch and the 60-inch Telescope at the Palomar Observatory as part of the Zwicky Transient Facility project. ZTF is supported by the National Science Foundation under Grant No. AST-1440341 and a collaboration including Caltech, IPAC, the Weizmann Institute for Science, the Oskar Klein Center at Stockholm University, the University of Maryland, the University of Washington, Deutsches Elektronen-Synchrotron and Humboldt University, Los Alamos National Laboratories, the TANGO Consortium of Taiwan, the University of Wisconsin at Milwaukee, and Lawrence Berkeley National Laboratories. Operations are conducted by COO, IPAC, and UW.

\bibliography{Bibliography}

\begin{thebibliography}{}

\bibitem[\protect\citeauthoryear{Alcock, Allsman, Alves, Axelrod, Bennett,
  Cook, Freeman, Griest, Guern, Lehner, et~al.}{Alcock
  et~al.}{1997}]{alcock1997macho}
Alcock, C., R.~Allsman, D.~Alves, T.~Axelrod, D.~Bennett, K.~Cook, K.~Freeman,
  K.~Griest, J.~Guern, M.~Lehner, et~al. (1997).
\newblock The macho project: 45 candidate microlensing events from the first
  year galactic bulge data.
\newblock {\em The Astrophysical Journal\/}~{\em 479\/}(1), 119.

\bibitem[\protect\citeauthoryear{Alcock, Allsman, Alves, Axelrod, Becker,
  Bennett, Cook, Drake, Freeman, Geha, et~al.}{Alcock
  et~al.}{2000}]{alcock2000macho}
Alcock, C., R.~Allsman, D.~R. Alves, T.~Axelrod, A.~C. Becker, D.~Bennett,
  K.~H. Cook, A.~J. Drake, K.~Freeman, M.~Geha, et~al. (2000).
\newblock The macho project: microlensing optical depth toward the galactic
  bulge from difference image analysis.
\newblock {\em The Astrophysical Journal\/}~{\em 541\/}(2), 734.

\bibitem[\protect\citeauthoryear{{Ansari}}{{Ansari}}{2004}]{ansari2004eros}
{Ansari}, R. (2004, July).
\newblock {EROS: a Galactic Microlensing Odyssey}.
\newblock {\em arXiv e-prints\/}, astro--ph/0407583.

\bibitem[\protect\citeauthoryear{Bachelet, Norbury, and Barclay}{Bachelet
  et~al.}{2017}]{ebachelet_2017_997468}
Bachelet, E., M.~Norbury, and T.~Barclay (2017, September).
\newblock ebachelet/pylima: pylima first release.

\bibitem[\protect\citeauthoryear{Batista, Keogh, Tataw, and De~Souza}{Batista
  et~al.}{2014}]{batista2014cid}
Batista, G.~E., E.~J. Keogh, O.~M. Tataw, and V.~M. De~Souza (2014).
\newblock Cid: an efficient complexity-invariant distance for time series.
\newblock {\em Data Mining and Knowledge Discovery\/}~{\em 28\/}(3), 634--669.

\bibitem[\protect\citeauthoryear{{Beaulieu}, {Bennett}, {Fouqu{\'e}},
  {Williams}, {Dominik}, {J{\o}rgensen}, {Kubas}, {Cassan}, {Coutures},
  {Greenhill}, {Hill}, {Menzies}, {Sackett}, {Albrow}, {Brillant}, {Caldwell},
  {Calitz}, {Cook}, {Corrales}, {Desort}, {Dieters}, {Dominis}, {Donatowicz},
  {Hoffman}, {Kane}, {Marquette}, {Martin}, {Meintjes}, {Pollard}, {Sahu},
  {Vinter}, {Wambsganss}, {Woller}, {Horne}, {Steele}, {Bramich}, {Burgdorf},
  {Snodgrass}, {Bode}, {Udalski}, {Szyma{\'n}ski}, {Kubiak}, {Wi{\c e}ckowski},
  {Pietrzy{\'n}ski}, {Soszy{\'n}ski}, {Szewczyk}, {Wyrzykowski},
  {Paczy{\'n}ski}, {Abe}, {Bond}, {Britton}, {Gilmore}, {Hearnshaw}, {Itow},
  {Kamiya}, {Kilmartin}, {Korpela}, {Masuda}, {Matsubara}, {Motomura},
  {Muraki}, {Nakamura}, {Okada}, {Ohnishi}, {Rattenbury}, {Sako}, {Sato},
  {Sasaki}, {Sekiguchi}, {Sullivan}, {Tristram}, {Yock}, and
  {Yoshioka}}{{Beaulieu} et~al.}{2006}]{microlensing_planet}
{Beaulieu}, J.-P., D.~P. {Bennett}, P.~{Fouqu{\'e}}, A.~{Williams},
  M.~{Dominik}, U.~G. {J{\o}rgensen}, D.~{Kubas}, A.~{Cassan}, C.~{Coutures},
  J.~{Greenhill}, K.~{Hill}, J.~{Menzies}, P.~D. {Sackett}, M.~{Albrow},
  S.~{Brillant}, J.~A.~R. {Caldwell}, J.~J. {Calitz}, K.~H. {Cook},
  E.~{Corrales}, M.~{Desort}, S.~{Dieters}, D.~{Dominis}, J.~{Donatowicz},
  M.~{Hoffman}, S.~{Kane}, J.-B. {Marquette}, R.~{Martin}, P.~{Meintjes},
  K.~{Pollard}, K.~{Sahu}, C.~{Vinter}, J.~{Wambsganss}, K.~{Woller},
  K.~{Horne}, I.~{Steele}, D.~M. {Bramich}, M.~{Burgdorf}, C.~{Snodgrass},
  M.~{Bode}, A.~{Udalski}, M.~K. {Szyma{\'n}ski}, M.~{Kubiak}, T.~{Wi{\c
  e}ckowski}, G.~{Pietrzy{\'n}ski}, I.~{Soszy{\'n}ski}, O.~{Szewczyk},
  {\L}.~{Wyrzykowski}, B.~{Paczy{\'n}ski}, F.~{Abe}, I.~A. {Bond}, T.~R.
  {Britton}, A.~C. {Gilmore}, J.~B. {Hearnshaw}, Y.~{Itow}, K.~{Kamiya}, P.~M.
  {Kilmartin}, A.~V. {Korpela}, K.~{Masuda}, Y.~{Matsubara}, M.~{Motomura},
  Y.~{Muraki}, S.~{Nakamura}, C.~{Okada}, K.~{Ohnishi}, N.~J. {Rattenbury},
  T.~{Sako}, S.~{Sato}, M.~{Sasaki}, T.~{Sekiguchi}, D.~J. {Sullivan}, P.~J.
  {Tristram}, P.~C.~M. {Yock}, and T.~{Yoshioka} (2006, January).
\newblock {Discovery of a cool planet of 5.5 Earth masses through gravitational
  microlensing}.
\newblock {\em Nat.\/}~{\em 439}, 437--440.

\bibitem[\protect\citeauthoryear{{Becker}, {Iben}, and {Tuggle}}{{Becker}
  et~al.}{1977}]{1977Becker}
{Becker}, S.~A., I.~{Iben}, Jr., and R.~S. {Tuggle} (1977, December).
\newblock {On the frequency-period distribution of Cepheid variables in
  galaxies in the Local Group}.
\newblock {\em ApJ\/}~{\em 218}, 633--653.

\bibitem[\protect\citeauthoryear{{Bellm}, {Kulkarni}, and {ZTF
  Collaboration}}{{Bellm} et~al.}{2015}]{BellmZTF}
{Bellm}, E.~C., S.~R. {Kulkarni}, and {ZTF Collaboration} (2015, January).
\newblock {The Zwicky Transient Facility}.
\newblock In {\em American Astronomical Society Meeting Abstracts \#225},
  Volume 225 of {\em American Astronomical Society Meeting Abstracts}, pp.\
  328.04.

\bibitem[\protect\citeauthoryear{Bennett, Anderson, Bond, Udalski, and
  Gould}{Bennett et~al.}{2006}]{bennett2006identification}
Bennett, D., J.~Anderson, I.~Bond, A.~Udalski, and A.~Gould (2006).
\newblock Identification of the ogle-2003-blg-235/moa-2003-blg-53 planetary
  host star.
\newblock {\em The Astrophysical Journal Letters\/}~{\em 647\/}(2), L171.

\bibitem[\protect\citeauthoryear{Bergstra and Bengio}{Bergstra and
  Bengio}{2012}]{bergstra2012random}
Bergstra, J. and Y.~Bengio (2012).
\newblock Random search for hyper-parameter optimization.
\newblock {\em Journal of Machine Learning Research\/}~{\em 13\/}(Feb),
  281--305.

\bibitem[\protect\citeauthoryear{Binney and Merrifield}{Binney and
  Merrifield}{1998}]{1998galactic_book}
Binney, J. and .~Merrifield, Michael (1998).
\newblock {\em Galactic astronomy}.
\newblock Princeton, NJ : Princeton University Press.
\newblock Includes bibliographical references (pages 745-776) and index.

\bibitem[\protect\citeauthoryear{{Bloom}, {Richards}, {Nugent}, {Quimby},
  {Kasliwal}, {Starr}, {Poznanski}, {Ofek}, {Cenko}, {Butler}, {Kulkarni},
  {Gal-Yam}, and {Law}}{{Bloom} et~al.}{2012}]{Bloom2012}
{Bloom}, J.~S., J.~W. {Richards}, P.~E. {Nugent}, R.~M. {Quimby}, M.~M.
  {Kasliwal}, D.~L. {Starr}, D.~{Poznanski}, E.~O. {Ofek}, S.~B. {Cenko}, N.~R.
  {Butler}, S.~R. {Kulkarni}, A.~{Gal-Yam}, and N.~{Law} (2012, November).
\newblock {Automating Discovery and Classification of Transients and Variable
  Stars in the Synoptic Survey Era}.
\newblock {\em PASP\/}~{\em 124}, 1175.

\bibitem[\protect\citeauthoryear{Bond, Abe, Dodd, Hearnshaw, Honda, Jugaku,
  Kilmartin, Marles, Masuda, Matsubara, et~al.}{Bond
  et~al.}{2001}]{bond2001real}
Bond, I., F.~Abe, R.~Dodd, J.~Hearnshaw, M.~Honda, J.~Jugaku, P.~Kilmartin,
  A.~Marles, K.~Masuda, Y.~Matsubara, et~al. (2001).
\newblock Real-time difference imaging analysis of moa galactic bulge
  observations during 2000.
\newblock {\em Monthly Notices of the Royal Astronomical Society\/}~{\em
  327\/}(3), 868--880.

\bibitem[\protect\citeauthoryear{Breiman}{Breiman}{1996}]{breiman1996bagging}
Breiman, L. (1996).
\newblock Bagging predictors.
\newblock {\em Machine learning\/}~{\em 24\/}(2), 123--140.

\bibitem[\protect\citeauthoryear{Breiman}{Breiman}{2001}]{Breiman2001}
Breiman, L. (2001, Oct).
\newblock Random forests.
\newblock {\em Machine Learning\/}~{\em 45\/}(1), 5--32.

\bibitem[\protect\citeauthoryear{Breiman, Friedman, Olshen, and Stone}{Breiman
  et~al.}{1984}]{breiman1984classification}
Breiman, L., J.~Friedman, R.~Olshen, and C.~Stone (1984).
\newblock Classification and regression trees. monterey, calif., usa:
  Wadsworth.

\bibitem[\protect\citeauthoryear{{Brink}, {Richards}, {Poznanski}, {Bloom},
  {Rice}, {Negahban}, and {Wainwright}}{{Brink} et~al.}{2013}]{Brink2013}
{Brink}, H., J.~W. {Richards}, D.~{Poznanski}, J.~S. {Bloom}, J.~{Rice},
  S.~{Negahban}, and M.~{Wainwright} (2013, October).
\newblock {Using machine learning for discovery in synoptic survey imaging
  data}.
\newblock {\em MNRAS\/}~{\em 435}, 1047--1060.

\bibitem[\protect\citeauthoryear{{Cao}, {Nugent}, and {Kasliwal}}{{Cao}
  et~al.}{2016}]{CaoIPTF}
{Cao}, Y., P.~E. {Nugent}, and M.~M. {Kasliwal} (2016, November).
\newblock {Intermediate Palomar Transient Factory: Realtime Image Subtraction
  Pipeline}.
\newblock {\em PASP\/}~{\em 128\/}(11), 114502.

\bibitem[\protect\citeauthoryear{Christ, Braun, Neuffer, and
  Kempa-Liehr}{Christ et~al.}{2018}]{Christ2018TimeSF}
Christ, M., N.~Braun, J.~Neuffer, and A.~W. Kempa-Liehr (2018).
\newblock Time series feature extraction on basis of scalable hypothesis tests
  (tsfresh - a python package).
\newblock {\em Neurocomputing\/}~{\em 307}, 72--77.

\bibitem[\protect\citeauthoryear{Cristianini and Shawe-Taylor}{Cristianini and
  Shawe-Taylor}{2000}]{cristianini2000introduction}
Cristianini, N. and J.~Shawe-Taylor (2000).
\newblock {\em An introduction to support vector machines and other
  kernel-based learning methods}.
\newblock Cambridge university press.

\bibitem[\protect\citeauthoryear{Cuillandre, Luppino, Starr, and
  Isani}{Cuillandre et~al.}{2000}]{cuillandre2000performance}
Cuillandre, J.-C., G.~A. Luppino, B.~M. Starr, and S.~Isani (2000).
\newblock Performance of the cfh12k: a 12k by 8k ccd mosaic camera for the cfht
  prime focus.
\newblock In {\em Optical and IR Telescope Instrumentation and Detectors},
  Volume 4008, pp.\  1010--1022. International Society for Optics and
  Photonics.

\bibitem[\protect\citeauthoryear{Djorgovski, Baltay, Mahabal, Drake, Williams,
  Rabinowitz, Graham, Donalek, Glikman, Bauer, et~al.}{Djorgovski
  et~al.}{2008}]{djorgovski2008palomar}
Djorgovski, S., C.~Baltay, A.~Mahabal, A.~Drake, R.~Williams, D.~Rabinowitz,
  M.~Graham, C.~Donalek, E.~Glikman, A.~Bauer, et~al. (2008).
\newblock The palomar-quest digital synoptic sky survey.
\newblock {\em Astronomische Nachrichten\/}~{\em 329\/}(3), 263--265.

\bibitem[\protect\citeauthoryear{{Gaia Collaboration}, {Brown}, {Vallenari},
  {Prusti}, {de Bruijne}, {Babusiaux}, {Bailer-Jones}, {Biermann}, {Evans},
  {Eyer}, and et~al.}{{Gaia Collaboration} et~al.}{2018}]{GAIA_DR2}
{Gaia Collaboration}, A.~G.~A. {Brown}, A.~{Vallenari}, T.~{Prusti}, J.~H.~J.
  {de Bruijne}, C.~{Babusiaux}, C.~A.~L. {Bailer-Jones}, M.~{Biermann}, D.~W.
  {Evans}, L.~{Eyer}, and et~al. (2018, August).
\newblock {Gaia Data Release 2. Summary of the contents and survey properties}.
\newblock {\em A\&A\/}~{\em 616}, A1.

\bibitem[\protect\citeauthoryear{Godines}{Godines}{2019}]{daniel_godines_2019_2541465}
Godines, D. (2019, January).
\newblock dgodinez77/lia: Version 1.0.

\bibitem[\protect\citeauthoryear{Griest}{Griest}{1991}]{griest1991galactic}
Griest, K. (1991).
\newblock Galactic microlensing as a method of detecting massive compact halo
  objects.
\newblock {\em The Astrophysical Journal\/}~{\em 366}, 412--421.

\bibitem[\protect\citeauthoryear{Han}{Han}{1999}]{han1999analytic}
Han, C. (1999).
\newblock Analytic relations between the observed gravitational microlensing
  parameters with and without the effect of blending.
\newblock {\em Monthly Notices of the Royal Astronomical Society\/}~{\em
  309\/}(2), 373--378.

\bibitem[\protect\citeauthoryear{Hellier}{Hellier}{2001}]{hellier2001cataclysmic}
Hellier, C. (2001).
\newblock {\em Cataclysmic Variable Stars - How and Why They Vary}.
\newblock Praxis Books in Astronomy and Space. Springer.

\bibitem[\protect\citeauthoryear{Hinton and Roweis}{Hinton and
  Roweis}{2003}]{hinton2003stochastic}
Hinton, G.~E. and S.~T. Roweis (2003).
\newblock Stochastic neighbor embedding.
\newblock In {\em Advances in neural information processing systems}, pp.\
  857--864.

\bibitem[\protect\citeauthoryear{{Hodgkin}, {Wyrzykowski}, {Blagorodnova}, and
  {Koposov}}{{Hodgkin} et~al.}{2013}]{Hodgkin2013}
{Hodgkin}, S.~T., L.~{Wyrzykowski}, N.~{Blagorodnova}, and S.~{Koposov} (2013,
  April).
\newblock {Transient astronomy with the Gaia satellite}.
\newblock {\em Philosophical Transactions of the Royal Society of London Series
  A\/}~{\em 371}, 20120239--20120239.

\bibitem[\protect\citeauthoryear{Howell, Nelson, and Rappaport}{Howell
  et~al.}{2001}]{howell2001exploration}
Howell, S.~B., L.~A. Nelson, and S.~Rappaport (2001).
\newblock An exploration of the paradigm for the 2-3 hour period gap in
  cataclysmic variables.
\newblock {\em The Astrophysical Journal\/}~{\em 550\/}(2), 897.

\bibitem[\protect\citeauthoryear{J.~Masci, R.~Laher, Rusholme, L.~Shupe, Groom,
  Surace, Jackson, Monkewitz, Beck, Flynn, Terek, Landry, Hacopians, Desai,
  Howell, Brooke, Imel, Wachter, Ye, and R.~Kulkarni}{J.~Masci
  et~al.}{2019}]{ZSDS_Masci}
J.~Masci, F., R.~R.~Laher, B.~Rusholme, D.~L.~Shupe, S.~Groom, J.~Surace,
  E.~Jackson, S.~Monkewitz, R.~Beck, D.~Flynn, S.~Terek, W.~Landry,
  E.~Hacopians, V.~Desai, J.~Howell, T.~Brooke, D.~Imel, S.~Wachter, Q.-Z. Ye,
  and S.~R.~Kulkarni (2019, 01).
\newblock The zwicky transient facility: Data processing, products, and
  archive.
\newblock {\em Publications of the Astronomical Society of the Pacific\/}~{\em
  131}, 018003.

\bibitem[\protect\citeauthoryear{Janczak, Fukui, Dong, Monard, Koz{\l}owski,
  Gould, Beaulieu, Kubas, Marquette, Sumi, et~al.}{Janczak
  et~al.}{2010}]{janczak2010sub}
Janczak, J., A.~Fukui, S.~Dong, L.~Monard, S.~Koz{\l}owski, A.~Gould,
  J.~Beaulieu, D.~Kubas, J.~Marquette, T.~Sumi, et~al. (2010).
\newblock Sub-saturn planet moa-2008-blg-310lb: likely to be in the galactic
  bulge.
\newblock {\em The Astrophysical Journal\/}~{\em 711\/}(2), 731.

\bibitem[\protect\citeauthoryear{Kim, Kim, Hwang, Albrow, Chung, Gould, Han,
  Jung, Ryu, Shin, et~al.}{Kim et~al.}{2018}]{kim2018korea}
Kim, D.-J., H.-W. Kim, K.-H. Hwang, M.~Albrow, S.-J. Chung, A.~Gould, C.~Han,
  Y.~Jung, Y.-H. Ryu, I.-G. Shin, et~al. (2018).
\newblock Korea microlensing telescope network microlensing events from 2015:
  Event-finding algorithm, vetting, and photometry.
\newblock {\em The Astronomical Journal\/}~{\em 155\/}(2), 76.

\bibitem[\protect\citeauthoryear{Kim, Lee, Park, Kim, Cha, Lee, Han, Chun, and
  Yuk}{Kim et~al.}{2016}]{kim2016kmtnet}
Kim, S.-L., C.-U. Lee, B.-G. Park, D.-J. Kim, S.-M. Cha, Y.~Lee, C.~Han, M.-Y.
  Chun, and I.~Yuk (2016).
\newblock Kmtnet: a network of 1.6 m wide-field optical telescopes installed at
  three southern observatories.
\newblock {\em Journal of the Korean Astronomical Society\/}~{\em 49\/}(1),
  37--44.

\bibitem[\protect\citeauthoryear{{Kochanek}, {Shappee}, {Stanek}, {Holoien},
  {Thompson}, {Prieto}, {Dong}, {Shields}, {Will}, {Britt}, {Perzanowski}, and
  {Pojma{\'n}ski}}{{Kochanek} et~al.}{2017}]{Kochanek2017}
{Kochanek}, C.~S., B.~J. {Shappee}, K.~Z. {Stanek}, T.~W.-S. {Holoien}, T.~A.
  {Thompson}, J.~L. {Prieto}, S.~{Dong}, J.~V. {Shields}, D.~{Will},
  C.~{Britt}, D.~{Perzanowski}, and G.~{Pojma{\'n}ski} (2017, October).
\newblock {The All-Sky Automated Survey for Supernovae (ASAS-SN) Light Curve
  Server v1.0}.
\newblock ~{\em 129\/}(10), 104502.

\bibitem[\protect\citeauthoryear{{Kulkarni}}{{Kulkarni}}{2013}]{Kulkarni}
{Kulkarni}, S.~R. (2013, February).
\newblock {The intermediate Palomar Transient Factory (iPTF) begins}.
\newblock {\em The Astronomer's Telegram\/}~{\em 4807}, 1.

\bibitem[\protect\citeauthoryear{{Law}, {Kulkarni}, {Dekany}, {Ofek}, {Quimby},
  {Nugent}, {Surace}, {Grillmair}, {Bloom}, {Kasliwal}, {Bildsten}, {Brown},
  {Cenko}, {Ciardi}, {Croner}, {Djorgovski}, {van Eyken}, {Filippenko}, {Fox},
  {Gal-Yam}, {Hale}, {Hamam}, {Helou}, {Henning}, {Howell}, {Jacobsen},
  {Laher}, {Mattingly}, {McKenna}, {Pickles}, {Poznanski}, {Rahmer}, {Rau},
  {Rosing}, {Shara}, {Smith}, {Starr}, {Sullivan}, {Velur}, {Walters}, and
  {Zolkower}}{{Law} et~al.}{2009}]{Law2009}
{Law}, N.~M., S.~R. {Kulkarni}, R.~G. {Dekany}, E.~O. {Ofek}, R.~M. {Quimby},
  P.~E. {Nugent}, J.~{Surace}, C.~C. {Grillmair}, J.~S. {Bloom}, M.~M.
  {Kasliwal}, L.~{Bildsten}, T.~{Brown}, S.~B. {Cenko}, D.~{Ciardi},
  E.~{Croner}, S.~G. {Djorgovski}, J.~{van Eyken}, A.~V. {Filippenko}, D.~B.
  {Fox}, A.~{Gal-Yam}, D.~{Hale}, N.~{Hamam}, G.~{Helou}, J.~{Henning}, D.~A.
  {Howell}, J.~{Jacobsen}, R.~{Laher}, S.~{Mattingly}, D.~{McKenna},
  A.~{Pickles}, D.~{Poznanski}, G.~{Rahmer}, A.~{Rau}, W.~{Rosing}, M.~{Shara},
  R.~{Smith}, D.~{Starr}, M.~{Sullivan}, V.~{Velur}, R.~{Walters}, and
  J.~{Zolkower} (2009, December).
\newblock {The Palomar Transient Factory: System Overview, Performance, and
  First Results}.
\newblock {\em PASP\/}~{\em 121}, 1395.

\bibitem[\protect\citeauthoryear{Maaten and Hinton}{Maaten and
  Hinton}{2008}]{maaten2008visualizing}
Maaten, L. v.~d. and G.~Hinton (2008).
\newblock Visualizing data using t-sne.
\newblock {\em Journal of machine learning research\/}~{\em 9\/}(Nov),
  2579--2605.

\bibitem[\protect\citeauthoryear{{Mao} and {Paczynski}}{{Mao} and
  {Paczynski}}{1991}]{binary_lens}
{Mao}, S. and B.~{Paczynski} (1991, June).
\newblock {Gravitational microlensing by double stars and planetary systems}.
\newblock {\em ApJl\/}~{\em 374}, L37--L40.

\bibitem[\protect\citeauthoryear{{Mislis}, {Bachelet}, {Alsubai}, {Bramich},
  and {Parley}}{{Mislis} et~al.}{2016}]{Mislis}
{Mislis}, D., E.~{Bachelet}, K.~A. {Alsubai}, D.~M. {Bramich}, and N.~{Parley}
  (2016, jan).
\newblock {SIDRA: a blind algorithm for signal detection in photometric
  surveys}.
\newblock {\em Monthly Notices of the RAS\/}~{\em 455}, 626--633.

\bibitem[\protect\citeauthoryear{{Narayan}, {Zaidi}, {Soraisam}, {Wang},
  {Lochner}, {Matheson}, {Saha}, {Yang}, {Zhao}, {Kececioglu}, {Scheidegger},
  {Snodgrass}, {Axelrod}, {Jenness}, {Maier}, {Ridgway}, {Seaman}, {Evans},
  {Singh}, {Taylor}, {Toeniskoetter}, {Welch}, {Zhu}, and {ANTARES
  Collaboration}}{{Narayan} et~al.}{2018}]{Gautham_Antares}
{Narayan}, G., T.~{Zaidi}, M.~D. {Soraisam}, Z.~{Wang}, M.~{Lochner},
  T.~{Matheson}, A.~{Saha}, S.~{Yang}, Z.~{Zhao}, J.~{Kececioglu},
  C.~{Scheidegger}, R.~T. {Snodgrass}, T.~{Axelrod}, T.~{Jenness}, R.~S.
  {Maier}, S.~T. {Ridgway}, R.~L. {Seaman}, E.~M. {Evans}, N.~{Singh},
  C.~{Taylor}, J.~{Toeniskoetter}, E.~{Welch}, S.~{Zhu}, and {ANTARES
  Collaboration} (2018, May).
\newblock {Machine-learning-based Brokers for Real-time Classification of the
  LSST Alert Stream}.
\newblock {\em The Astrophysical Journal Supplement Series\/}~{\em 236}, 9.

\bibitem[\protect\citeauthoryear{{Odewahn}, {de Carvalho}, {Gal}, {Djorgovski},
  {Brunner}, {Mahabal}, {Lopes}, {Moreira}, and {Stalder}}{{Odewahn}
  et~al.}{2004}]{Odewahn2004}
{Odewahn}, S.~C., R.~R. {de Carvalho}, R.~R. {Gal}, S.~G. {Djorgovski},
  R.~{Brunner}, A.~{Mahabal}, P.~A.~A. {Lopes}, J.~L.~K. {Moreira}, and
  B.~{Stalder} (2004, December).
\newblock {The Digitized Second Palomar Observatory Sky Survey (DPOSS). III.
  Star-Galaxy Separation}.
\newblock {\em AJ\/}~{\em 128}, 3092--3107.

\bibitem[\protect\citeauthoryear{Osaki}{Osaki}{1974}]{osaki1974accretion}
Osaki, Y. (1974).
\newblock An accretion model for the outbursts of u geminorum stars.
\newblock {\em Publications of the Astronomical Society of Japan\/}~{\em 26},
  429--436.

\bibitem[\protect\citeauthoryear{Paczynski}{Paczynski}{1986}]{paczynski1986gravitational}
Paczynski, B. (1986).
\newblock Gravitational microlensing by the galactic halo.
\newblock {\em The Astrophysical Journal\/}~{\em 304}, 1--5.

\bibitem[\protect\citeauthoryear{{Pashchenko}, {Sokolovsky}, and
  {Gavras}}{{Pashchenko} et~al.}{2018}]{Pashchenko2018}
{Pashchenko}, I.~N., K.~V. {Sokolovsky}, and P.~{Gavras} (2018, April).
\newblock {Machine learning search for variable stars}.
\newblock {\em MNRAS\/}~{\em 475}, 2326--2343.

\bibitem[\protect\citeauthoryear{{Pearson}}{{Pearson}}{1901}]{Pearson_PCA}
{Pearson}, K. (1901).
\newblock Liii. on lines and planes of closest fit to systems of points in
  space.
\newblock {\em The London, Edinburgh, and Dublin Philosophical Magazine and
  Journal of Science\/}~{\em 2\/}(11), 559--572.

\bibitem[\protect\citeauthoryear{Pedregosa, Varoquaux, Gramfort, Michel,
  Thirion, Grisel, Blondel, Prettenhofer, Weiss, Dubourg, Vanderplas, Passos,
  Cournapeau, Brucher, Perrot, and Duchesnay}{Pedregosa
  et~al.}{2011}]{scikit-learn}
Pedregosa, F., G.~Varoquaux, A.~Gramfort, V.~Michel, B.~Thirion, O.~Grisel,
  M.~Blondel, P.~Prettenhofer, R.~Weiss, V.~Dubourg, J.~Vanderplas, A.~Passos,
  D.~Cournapeau, M.~Brucher, M.~Perrot, and E.~Duchesnay (2011).
\newblock {Scikit-learn: Machine Learning in Python }.
\newblock {\em Journal of Machine Learning Research\/}~{\em 12}, 2825--2830.

\bibitem[\protect\citeauthoryear{Price-Whelan, Agueros, Fournier, Street, Ofek,
  Covey, Levitan, Laher, Sesar, and Surace}{Price-Whelan
  et~al.}{2014}]{price2014statistical}
Price-Whelan, A., M.~Agueros, A.~Fournier, R.~Street, E.~Ofek, K.~Covey,
  D.~Levitan, R.~Laher, B.~Sesar, and J.~Surace (2014).
\newblock Statistical searches for microlensing events in large, non-uniformly
  sampled time-domain surveys: A test using palomar transient factory data.
\newblock {\em The Astrophysical Journal\/}~{\em 781\/}(1), 35.

\bibitem[\protect\citeauthoryear{Rau, Kulkarni, Law, Bloom, Ciardi, Djorgovski,
  Fox, Gal-Yam, Grillmair, Kasliwal, et~al.}{Rau
  et~al.}{2009}]{rau2009exploring}
Rau, A., S.~R. Kulkarni, N.~M. Law, J.~S. Bloom, D.~Ciardi, G.~S. Djorgovski,
  D.~B. Fox, A.~Gal-Yam, C.~C. Grillmair, M.~M. Kasliwal, et~al. (2009).
\newblock Exploring the optical transient sky with the palomar transient
  factory.
\newblock {\em Publications of the Astronomical Society of the Pacific\/}~{\em
  121\/}(886), 1334.

\bibitem[\protect\citeauthoryear{{Richards}, {Starr}, {Butler}, {Bloom},
  {Brewer}, {Crellin-Quick}, {Higgins}, {Kennedy}, and {Rischard}}{{Richards}
  et~al.}{2011}]{Richards2011}
{Richards}, J.~W., D.~L. {Starr}, N.~R. {Butler}, J.~S. {Bloom}, J.~M.
  {Brewer}, A.~{Crellin-Quick}, J.~{Higgins}, R.~{Kennedy}, and M.~{Rischard}
  (2011, May).
\newblock {On Machine-learned Classification of Variable Stars with Sparse and
  Noisy Time-series Data}.
\newblock {\em ApJ\/}~{\em 733}, 10.

\bibitem[\protect\citeauthoryear{Richman and Moorman}{Richman and
  Moorman}{2000}]{richman2000physiological}
Richman, J.~S. and J.~R. Moorman (2000).
\newblock Physiological time-series analysis using approximate entropy and
  sample entropy.
\newblock {\em American Journal of Physiology-Heart and Circulatory
  Physiology\/}~{\em 278\/}(6), H2039--H2049.

\bibitem[\protect\citeauthoryear{Ripley}{Ripley}{1996}]{ripley1996pattern}
Ripley, B.~D. (1996).
\newblock Pattern recognition via neural networks.
\newblock {\em a volume of Oxford Graduate Lectures on Neural Networks, title
  to be decided. Oxford University Press.[See http://www. stats. ox. ac.
  uk/ripley/papers. html.]\/}.

\bibitem[\protect\citeauthoryear{Robinson}{Robinson}{1976}]{robinson1976structure}
Robinson, E.~L. (1976).
\newblock The structure of cataclysmic variables.
\newblock {\em Annual review of astronomy and astrophysics\/}~{\em 14\/}(1),
  119--142.

\bibitem[\protect\citeauthoryear{Romano, Aragon, and Ding}{Romano
  et~al.}{2006}]{romano2006supernova}
Romano, R.~A., C.~R. Aragon, and C.~Ding (2006).
\newblock Supernova recognition using support vector machines.
\newblock In {\em Machine Learning and Applications, 2006. ICMLA'06. 5th
  International Conference on}, pp.\  77--82. IEEE.

\bibitem[\protect\citeauthoryear{Schreiber and Schmitz}{Schreiber and
  Schmitz}{1997}]{schreiber1997discrimination}
Schreiber, T. and A.~Schmitz (1997).
\newblock Discrimination power of measures for nonlinearity in a time series.
\newblock {\em Physical Review E\/}~{\em 55\/}(5), 5443.

\bibitem[\protect\citeauthoryear{Schreiber and Schmitz}{Schreiber and
  Schmitz}{2000}]{schreiber2000surrogate}
Schreiber, T. and A.~Schmitz (2000).
\newblock Surrogate time series.
\newblock {\em Physica D: Nonlinear Phenomena\/}~{\em 142\/}(3-4), 346--382.

\bibitem[\protect\citeauthoryear{{Sesar}, {Ivezi{\'c}}, {Grammer}, {Morgan},
  {Becker}, {Juri{\'c}}, {De Lee}, {Annis}, {Beers}, {Fan}, {Lupton}, {Gunn},
  {Knapp}, {Jiang}, {Jester}, {Johnston}, and {Lampeitl}}{{Sesar}
  et~al.}{2010}]{2010Sesar}
{Sesar}, B., {\v Z}.~{Ivezi{\'c}}, S.~H. {Grammer}, D.~P. {Morgan}, A.~C.
  {Becker}, M.~{Juri{\'c}}, N.~{De Lee}, J.~{Annis}, T.~C. {Beers}, X.~{Fan},
  R.~H. {Lupton}, J.~E. {Gunn}, G.~R. {Knapp}, L.~{Jiang}, S.~{Jester}, D.~E.
  {Johnston}, and H.~{Lampeitl} (2010, January).
\newblock {Light Curve Templates and Galactic Distribution of RR Lyrae Stars
  from Sloan Digital Sky Survey Stripe 82}.
\newblock {\em ApJ\/}~{\em 708}, 717--741.

\bibitem[\protect\citeauthoryear{Shannon and Weaver}{Shannon and
  Weaver}{1949}]{shannon1949mathematical}
Shannon, C. and W.~Weaver (1949).
\newblock The mathematical study of communication.
\newblock {\em Urbana, IL: University of Illinois Press\/}.

\bibitem[\protect\citeauthoryear{{Shappee}, {Prieto}, {Grupe}, {Kochanek},
  {Stanek}, {De Rosa}, {Mathur}, {Zu}, {Peterson}, {Pogge}, {Komossa}, {Im},
  {Jencson}, {Holoien}, {Basu}, {Beacom}, {Szczygie{\l}}, {Brimacombe},
  {Adams}, {Campillay}, {Choi}, {Contreras}, {Dietrich}, {Dubberley},
  {Elphick}, {Foale}, {Giustini}, {Gonzalez}, {Hawkins}, {Howell}, {Hsiao},
  {Koss}, {Leighly}, {Morrell}, {Mudd}, {Mullins}, {Nugent}, {Parrent},
  {Phillips}, {Pojmanski}, {Rosing}, {Ross}, {Sand}, {Terndrup}, {Valenti},
  {Walker}, and {Yoon}}{{Shappee} et~al.}{2014}]{Shappee2014}
{Shappee}, B.~J., J.~L. {Prieto}, D.~{Grupe}, C.~S. {Kochanek}, K.~Z. {Stanek},
  G.~{De Rosa}, S.~{Mathur}, Y.~{Zu}, B.~M. {Peterson}, R.~W. {Pogge},
  S.~{Komossa}, M.~{Im}, J.~{Jencson}, T.~W.-S. {Holoien}, U.~{Basu}, J.~F.
  {Beacom}, D.~M. {Szczygie{\l}}, J.~{Brimacombe}, S.~{Adams}, A.~{Campillay},
  C.~{Choi}, C.~{Contreras}, M.~{Dietrich}, M.~{Dubberley}, M.~{Elphick},
  S.~{Foale}, M.~{Giustini}, C.~{Gonzalez}, E.~{Hawkins}, D.~A. {Howell}, E.~Y.
  {Hsiao}, M.~{Koss}, K.~M. {Leighly}, N.~{Morrell}, D.~{Mudd}, D.~{Mullins},
  J.~M. {Nugent}, J.~{Parrent}, M.~M. {Phillips}, G.~{Pojmanski}, W.~{Rosing},
  R.~{Ross}, D.~{Sand}, D.~M. {Terndrup}, S.~{Valenti}, Z.~{Walker}, and
  Y.~{Yoon} (2014, June).
\newblock {The Man behind the Curtain: X-Rays Drive the UV through NIR
  Variability in the 2013 Active Galactic Nucleus Outburst in NGC 2617}.
\newblock ~{\em 788}, 48.

\bibitem[\protect\citeauthoryear{Shin, Sekora, and Byun}{Shin
  et~al.}{2009}]{shin2009detecting}
Shin, M.-S., M.~Sekora, and Y.-I. Byun (2009).
\newblock Detecting variability in massive astronomical time series data--i.
  application of an infinite gaussian mixture model.
\newblock {\em Monthly Notices of the Royal Astronomical Society\/}~{\em
  400\/}(4), 1897--1910.

\bibitem[\protect\citeauthoryear{{Shlens}}{{Shlens}}{2014}]{2014PCA}
{Shlens}, J. (2014, April).
\newblock {A Tutorial on Principal Component Analysis}.
\newblock {\em arXiv e-prints\/}, arXiv:1404.1100.

\bibitem[\protect\citeauthoryear{Shvartzvald, Bryden, Gould, Henderson, Howell,
  and Beichman}{Shvartzvald et~al.}{2017}]{shvartzvald2017ukirt}
Shvartzvald, Y., G.~Bryden, A.~Gould, C.~Henderson, S.~Howell, and C.~Beichman
  (2017).
\newblock Ukirt microlensing surveys as a pathfinder for wfirst: The detection
  of five highly extinguished low-events.
\newblock {\em The Astronomical Journal\/}~{\em 153\/}(2), 61.

\bibitem[\protect\citeauthoryear{Stetson}{Stetson}{1996}]{stetson1996automatic}
Stetson, P. (1996).
\newblock On the automatic determination of light-curve parameters for cepheid
  variables.
\newblock {\em Publications of the Astronomical Society of the Pacific\/}~{\em
  108\/}(728), 851.

\bibitem[\protect\citeauthoryear{Sumi and Penny}{Sumi and
  Penny}{2016}]{sumi2016possible}
Sumi, T. and M.~Penny (2016).
\newblock Possible solution of the long-standing discrepancy in the
  microlensing optical depth toward the galactic bulge by correcting the
  stellar number count.
\newblock {\em The Astrophysical Journal\/}~{\em 827\/}(2), 139.

\bibitem[\protect\citeauthoryear{Sutherland, Emerson, Dalton, Atad-Ettedgui,
  Beard, Bennett, Bezawada, Born, Caldwell, Clark, et~al.}{Sutherland
  et~al.}{2015}]{VISTA_2015}
Sutherland, W., J.~Emerson, G.~Dalton, E.~Atad-Ettedgui, S.~Beard, R.~Bennett,
  N.~Bezawada, A.~Born, M.~Caldwell, P.~Clark, et~al. (2015).
\newblock The visible and infrared survey telescope for astronomy (vista):
  design, technical overview, and performance.
\newblock {\em Astronomy \& Astrophysics\/}~{\em 575}, A25.

\bibitem[\protect\citeauthoryear{Tsapras}{Tsapras}{2018}]{tsapras2018microlensing}
Tsapras, Y. (2018).
\newblock Microlensing searches for exoplanets.
\newblock {\em Geosciences\/}~{\em 8\/}(10), 365.

\bibitem[\protect\citeauthoryear{Tsapras, Hundertmark, Horne, Udalski,
  Snodgrass, Street, Bramich, Dominik, Bozza, Jaimes, et~al.}{Tsapras
  et~al.}{2016}]{tsapras2016ogle}
Tsapras, Y., M.~Hundertmark, K.~Horne, A.~Udalski, C.~Snodgrass, R.~Street,
  D.~Bramich, M.~Dominik, V.~Bozza, R.~Jaimes, et~al. (2016).
\newblock The ogle-iii planet detection efficiency from six years of
  microlensing observations (2003--2008).
\newblock {\em Monthly Notices of the Royal Astronomical Society\/}~{\em
  457\/}(2), 1320--1331.

\bibitem[\protect\citeauthoryear{Tyson}{Tyson}{2002}]{tyson2002_lsst}
Tyson, J.~A. (2002).
\newblock Large synoptic survey telescope: overview.
\newblock In {\em Survey and Other Telescope Technologies and Discoveries},
  Volume 4836, pp.\  10--21. International Society for Optics and Photonics.

\bibitem[\protect\citeauthoryear{{Udalski}}{{Udalski}}{2003}]{udalski2003optical}
{Udalski}, A. (2003, December).
\newblock {The Optical Gravitational Lensing Experiment. Real Time Data
  Analysis Systems in the OGLE-III Survey}.
\newblock {\em ACTAA\/}~{\em 53}, 291--305.

\bibitem[\protect\citeauthoryear{{Udalski}, {Kubiak}, and
  {Szymanski}}{{Udalski} et~al.}{1997}]{1997_OGLEII}
{Udalski}, A., M.~{Kubiak}, and M.~{Szymanski} (1997, July).
\newblock {Optical Gravitational Lensing Experiment. OGLE-2 -- the Second Phase
  of the OGLE Project}.
\newblock {\em Acta Astron.\/}~{\em 47}, 319--344.

\bibitem[\protect\citeauthoryear{Udalski, Szymanski, Kaluzny, Kubiak, and
  Mateo}{Udalski et~al.}{1992}]{udalski1992optical}
Udalski, A., M.~Szymanski, J.~Kaluzny, M.~Kubiak, and M.~Mateo (1992).
\newblock The optical gravitational lensing experiment.
\newblock {\em Acta Astronomica\/}~{\em 42}, 253--284.

\bibitem[\protect\citeauthoryear{{Udalski}, {Szyma{\'n}ski}, and
  {Szyma{\'n}ski}}{{Udalski} et~al.}{2015}]{2015OGLE_4}
{Udalski}, A., M.~K. {Szyma{\'n}ski}, and G.~{Szyma{\'n}ski} (2015, March).
\newblock {OGLE-IV: Fourth Phase of the Optical Gravitational Lensing
  Experiment}.
\newblock {\em Acta Astron.\/}~{\em 65}, 1--38.

\bibitem[\protect\citeauthoryear{{van Velzen}, {Gezari}, {Cenko}, {Kara},
  {Miller-Jones}, {Hung}, {Bright}, {Roth}, {Blagorodnova}, {Huppenkothen},
  {Yan}, {Ofek}, {Sollerman}, {Frederick}, {Ward}, {Graham}, {Fender},
  {Kasliwal}, {Canella}, {Stein}, {Giomi}, {Brinnel}, {Santen}, {Nordin},
  {Bellm}, {Dekany}, {Fremling}, {Golkhou}, {Kupfer}, {Kulkarni}, {Laher},
  {Mahabal}, {Masci}, {Miller}, {Neill}, {Riddle}, {Rigault}, {Rusholme},
  {Soumagnac}, and {Tachibana}}{{van Velzen} et~al.}{2018}]{2018_ztf_tde}
{van Velzen}, S., S.~{Gezari}, S.~{Cenko}, E.~{Kara}, J.~C. {Miller-Jones},
  T.~{Hung}, J.~{Bright}, N.~{Roth}, N.~{Blagorodnova}, D.~{Huppenkothen},
  L.~{Yan}, E.~{Ofek}, J.~{Sollerman}, S.~{Frederick}, C.~{Ward}, M.~J.
  {Graham}, R.~{Fender}, M.~M. {Kasliwal}, C.~{Canella}, R.~{Stein},
  M.~{Giomi}, V.~{Brinnel}, J.~{Santen}, J.~{Nordin}, E.~C. {Bellm},
  R.~{Dekany}, C.~{Fremling}, V.~{Golkhou}, T.~{Kupfer}, S.~R. {Kulkarni},
  R.~R. {Laher}, A.~{Mahabal}, F.~J. {Masci}, A.~A. {Miller}, J.~D. {Neill},
  R.~{Riddle}, M.~{Rigault}, B.~{Rusholme}, M.~T. {Soumagnac}, and
  Y.~{Tachibana} (2018, September).
\newblock {The first tidal disruption flare in ZTF: from photometric selection
  to multi-wavelength characterization}.
\newblock {\em arXiv e-prints\/}, arXiv:1809.02608.

\bibitem[\protect\citeauthoryear{Vanderplas}{Vanderplas}{2015}]{jake_vanderplas_2015_14833}
Vanderplas, J. (2015, February).
\newblock {gatspy: General tools for Astronomical Time Series in Python}.

\bibitem[\protect\citeauthoryear{{VanderPlas} and {Ivezi{\'c}}}{{VanderPlas}
  and {Ivezi{\'c}}}{2015}]{2015gatspy}
{VanderPlas}, J.~T. and {\v Z}.~{Ivezi{\'c}} (2015, October).
\newblock {Periodograms for Multiband Astronomical Time Series}.
\newblock {\em ApJ\/}~{\em 812}, 18.

\bibitem[\protect\citeauthoryear{Von~Neumann}{Von~Neumann}{1941}]{von1941distribution}
Von~Neumann, J. (1941).
\newblock Distribution of the ratio of the mean square successive difference to
  the variance.
\newblock {\em The Annals of Mathematical Statistics\/}~{\em 12\/}(4),
  367--395.

\bibitem[\protect\citeauthoryear{{Weir}, {Fayyad}, {Djorgovski}, and
  {Roden}}{{Weir} et~al.}{1995}]{Weir1995}
{Weir}, N., U.~M. {Fayyad}, S.~G. {Djorgovski}, and J.~{Roden} (1995,
  December).
\newblock {The SKICAT System for Processing and Analyzing Digital Imaging Sky
  Surveys}.
\newblock {\em PASP\/}~{\em 107}, 1243.

\bibitem[\protect\citeauthoryear{{Wozniak}}{{Wozniak}}{2000}]{2000Wozniak}
{Wozniak}, P.~R. (2000, December).
\newblock {Difference Image Analysis of the OGLE-II Bulge Data. I. The Method}.
\newblock {\em Acta Astron.\/}~{\em 50}, 421--450.

\bibitem[\protect\citeauthoryear{Wyrzykowski, Kostrzewa-Rutkowska, Skowron,
  Rybicki, Mr{\'o}z, Koz{\l}owski, Udalski, Szyma{\'n}ski, Pietrzy{\'n}ski,
  Soszy{\'n}ski, et~al.}{Wyrzykowski et~al.}{2016}]{wyrzykowski2016black}
Wyrzykowski, {\L}., Z.~Kostrzewa-Rutkowska, J.~Skowron, K.~Rybicki,
  P.~Mr{\'o}z, S.~Koz{\l}owski, A.~Udalski, M.~Szyma{\'n}ski,
  G.~Pietrzy{\'n}ski, I.~Soszy{\'n}ski, et~al. (2016).
\newblock Black hole, neutron star and white dwarf candidates from microlensing
  with ogle-iii.
\newblock {\em Monthly Notices of the Royal Astronomical Society\/}~{\em
  458\/}(3), 3012--3026.

\bibitem[\protect\citeauthoryear{{Wyrzykowski}, {Kostrzewa-Rutkowska},
  {Skowron}, {Rybicki}, {Mr{\'o}z}, {Koz{\l}owski}, {Udalski}, {Szyma{\'n}ski},
  {Pietrzy{\'n}ski}, {Soszy{\'n}ski}, {Ulaczyk}, {Pietrukowicz}, {Poleski},
  {Pawlak}, {I{\l}kiewicz}, and {Rattenbury}}{{Wyrzykowski}
  et~al.}{2016}]{2016_OGLE_RF}
{Wyrzykowski}, {\L}., Z.~{Kostrzewa-Rutkowska}, J.~{Skowron}, K.~A. {Rybicki},
  P.~{Mr{\'o}z}, S.~{Koz{\l}owski}, A.~{Udalski}, M.~K. {Szyma{\'n}ski},
  G.~{Pietrzy{\'n}ski}, I.~{Soszy{\'n}ski}, K.~{Ulaczyk}, P.~{Pietrukowicz},
  R.~{Poleski}, M.~{Pawlak}, K.~{I{\l}kiewicz}, and N.~J. {Rattenbury} (2016,
  May).
\newblock {Black hole, neutron star and white dwarf candidates from
  microlensing with OGLE-III}.
\newblock {\em MNRAS\/}~{\em 458}, 3012--3026.

\bibitem[\protect\citeauthoryear{{Wyrzykowski}, {Rynkiewicz}, {Skowron},
  {Koz{\l}owski}, {Udalski}, {Szyma{\'n}ski}, {Kubiak}, {Soszy{\'n}ski},
  {Pietrzy{\'n}ski}, {Poleski}, {Pietrukowicz}, and {Pawlak}}{{Wyrzykowski}
  et~al.}{2015a}]{2015_OGLE_RF}
{Wyrzykowski}, {\L}., A.~E. {Rynkiewicz}, J.~{Skowron}, S.~{Koz{\l}owski},
  A.~{Udalski}, M.~K. {Szyma{\'n}ski}, M.~{Kubiak}, I.~{Soszy{\'n}ski},
  G.~{Pietrzy{\'n}ski}, R.~{Poleski}, P.~{Pietrukowicz}, and M.~{Pawlak}
  (2015a, January).
\newblock {OGLE-III Microlensing Events and the Structure of the Galactic
  Bulge}.
\newblock ~{\em 216}, 12.

\bibitem[\protect\citeauthoryear{{Wyrzykowski}, {Rynkiewicz}, {Skowron},
  {Koz{\l}owski}, {Udalski}, {Szyma{\'n}ski}, {Kubiak}, {Soszy{\'n}ski},
  {Pietrzy{\'n}ski}, {Poleski}, {Pietrukowicz}, and {Pawlak}}{{Wyrzykowski}
  et~al.}{2015b}]{2015_Wyrzykowski}
{Wyrzykowski}, {\L}., A.~E. {Rynkiewicz}, J.~{Skowron}, S.~{Koz{\l}owski},
  A.~{Udalski}, M.~K. {Szyma{\'n}ski}, M.~{Kubiak}, I.~{Soszy{\'n}ski},
  G.~{Pietrzy{\'n}ski}, R.~{Poleski}, P.~{Pietrukowicz}, and M.~{Pawlak}
  (2015b, January).
\newblock {OGLE-III Microlensing Events and the Structure of the Galactic
  Bulge}.
\newblock ~{\em 216}, 12.

\bibitem[\protect\citeauthoryear{{Wyrzykowski}, {Udalski}, {Kubiak},
  {Szymanski}, {Zebrun}, {Soszynski}, {Wozniak}, {Pietrzynski}, and
  {Szewczyk}}{{Wyrzykowski} et~al.}{2003}]{OGLE_MachineLEarning2003}
{Wyrzykowski}, L., A.~{Udalski}, M.~{Kubiak}, M.~{Szymanski}, K.~{Zebrun},
  I.~{Soszynski}, P.~R. {Wozniak}, G.~{Pietrzynski}, and O.~{Szewczyk} (2003,
  March).
\newblock {The Optical Gravitational Lensing Experiment. Eclipsing Binary Stars
  in the Large Magellanic Cloud}.
\newblock {\em ACTAA\/}~{\em 53}, 1--25.

\end{thebibliography}
\end{document}